\documentclass[twocolumn,epjc3]{svjour3}

\RequirePackage{fix-cm}
\RequirePackage{graphicx}
\RequirePackage{bm}
\RequirePackage{amsmath,amsfonts,amssymb,amscd}
\RequirePackage{slashed}
\RequirePackage{physics}
\RequirePackage{microtype}

\journalname{Eur. Phys. J. C}

\begin{document}

\title{Heavy Neutrinos across the Electroweak-to-Multi-TeV Frontier via Novel ML--Enhanced Probes}

\author{Yin-Fa Shen\thanksref{e1, addr1}
        \and
        Alfredo Gurrola\thanksref{e2,addr1}
        \and
        Francesco Romeo\thanksref{e3,addr1}
        \and
        Denis Rathjens\thanksref{e4,addr1}
        \and
        Andr\'es Fl\'orez\thanksref{e5,addr2}
}

\thankstext{e1}{e-mail: yin-fa.shen@Vanderbilt.Edu}
\thankstext{e2}{e-mail: alfredo.gurrola@Vanderbilt.Edu}
\thankstext{e3}{e-mail: francesco.romeo@cern.ch}
\thankstext{e4}{e-mail: denis.rathjens@cern.ch}
\thankstext{e5}{e-mail: ca.florez@uniandes.edu.co}

\institute{Department of Physics and Astronomy, Vanderbilt University, Nashville, TN, 37235, USA. \label{addr1}\and 
Department of Physics, Universidad de los Andes,
Bogotá, Colombia. \label{addr2}
}

\date{Received: 4 February 2026 / Accepted: 21 October 2026}

\maketitle

\begin{abstract}
We propose a new strategy to probe heavy neutrinos with non-universal fermion couplings at the Large Hadron Collider (LHC) using a novel production mechanism and machine-learning algorithms. Focusing on proton--proton collisions at $\sqrt{s} = 13.6~\mathrm{TeV}$, we investigate final states containing a charged lepton, missing transverse energy, and two jets. For heavy neutrino masses below $\order{1~\mathrm{TeV}}$, production is dominated by the $s$ channel process. At higher masses, vector boson fusion becomes the dominant production mechanism, with cross sections that decrease slowly as the heavy neutrino mass increases. We simulate both signal and Standard Model background events and employ gradient-boosted decision trees to optimize event classification. Assuming an integrated luminosity of $3000~\mathrm{fb^{-1}}$, expected for the high-luminosity, and considering realistic statistical and systematic uncertainties, we find that heavy neutrinos in the mass range $50~\mathrm{GeV}$--$10~\mathrm{TeV}$ can be probed with sensitivity to the mixing parameter $\abs{V_{\ell N}}^2$ spanning from $\order{10^{-5}}$ to 1. This approach enhances the discovery potential for heavy neutrinos and provides a complementary pathway to existing search strategies.
\end{abstract}

\section{Introduction}

Searching for new physics beyond the Standard Model (SM) remains a central goal in particle physics. Compelling evidence for physics beyond the minimal SM already exists in the neutrino sector--namely, neutrino oscillations~\cite{Super-Kamiokande:1998kpq,SNO:2002tuh,DayaBay:2012fng}, which imply nonzero but tiny neutrino masses~\cite{Mohapatra:2006gs}. While such masses could, in principle, arise from Yukawa couplings analogous to those in the quark sector, the required couplings would need to be unnaturally small~\cite{Mohapatra:2006gs}. To address this issue, various theoretical frameworks have been proposed to naturally explain the smallness of neutrino masses~\cite{Minkowski:1977sc,Gell-Mann:1979vob,Yanagida:1979as,Schechter:1980gr,Magg:1980ut,Cheng:1980qt,Lazarides:1980nt,Mohapatra:1980yp,Foot:1988aq,Arkani-Hamed:1998wuz}. A common feature among many of these models is the presence of heavy Majorana neutrinos, which generate light neutrino masses inversely proportional to a large mass scale through the seesaw mechanism~\cite{Minkowski:1977sc,Gell-Mann:1979vob,Yanagida:1979as}. A key phenomenological prediction of this mechanism is the existence of heavy neutrinos. Therefore, searching for such heavy states is essential for testing the seesaw mechanism and potentially uncovering the origin of neutrino masses.

In seesaw-inspired models, the masses of heavy neutrinos can span a vast range, from the eV scale up to $10^{16}~\mathrm{GeV}$~\cite{Drewes:2015jna}. For heavy neutrinos lighter than $\order{1~\mathrm{GeV}}$, constraints arise from meson decays~\cite{T2K:2019jwa,NA62:2020mcv,BESIII:2019oef,E949:2014gsn,PIENU:2017wbj,LHCb:2014osd,Belle:2013ytx} and neutrinoless double-beta decay ($0\nu\beta\beta$)~\cite{GERDA:2020xhi}. At higher masses, electroweak precision data (EWPD) provide useful limits~\cite{Abada:2014cca,Abada:2015zea}, while extensive direct searches have been carried out at LEP and the LHC by experiments such as ALEPH, DELPHI, CMS, and ATLAS~\cite{ALEPH:1991qhf,DELPHI:1996qcc,CMS:2018iaf,CMS:2018jxx,CMS:2022hvh,CMS:2022fut,CMS:2023jqi,CMS:2024xdq,ATLAS:2019kpx,ATLAS:2023tkz,ATLAS:2024rzi,ATLAS:2024fcs,CMS:2024hik,CMS:2024ake,ATLAS:2025uah}.

For masses up to $\order{10^2~\mathrm{GeV}}$, considerable effort has been devoted to assessing the discovery prospects at future lepton colliders such as CEPC and FCC-ee~\cite{Fargion:1995qb,Fargion:1999ss,Antusch:2015mia,Antusch:2016ejd,Antusch:2016vyf,Liao:2017jiz,Wang:2019xvx,Ding:2019tqq,Gao:2021one,Blondel:2021ema,Knapen:2021svn,Shen:2022ffi,Bellagamba:2025xpd,Ruiz:2017yyf}. At even higher masses, the LHC remains the primary facility for probing heavy neutrinos: in the sub-TeV region, they can be resonantly produced and their decay products fully reconstructed, whereas at larger masses the production cross section becomes increasingly suppressed by phase space~\cite{Alva:2014gxa,Das:2015toa,Das:2016hof,Das:2017gke,Florez:2017xhf}. In this regime, $t$ channel mechanisms such as vector boson fusion (VBF) offer a more robust alternative, as their cross section degrades more slowly with the heavy neutrino mass~\cite{Delannoy:2013ata,CMS:2016ucr,Fuks:2020att,CMS:2024vhy}. As a result, an inclusive simulation and search strategy for the process $pp \to \mu \nu jj$, which receives contributions from both resonant and $t$ channel diagrams, ensures coverage over a wide range of heavy neutrino masses.

In this work, we analyze and study simulated signal and background samples to develop a new novel strategy to probe heavy neutrinos in proton--proton ($pp$) collisions at the LHC with a center-of-mass energy of $\sqrt{s} = 13.6~\mathrm{TeV}$. We focus on the final state $\ell \nu jj$, where $\ell$ and $\nu$ denote a charged lepton and a (light) neutrino, respectively, and $jj$ denotes two jets. Representative Feynman diagrams contributing to signal processes are shown in Fig.~\ref{fig: Feyn}. Owing to the nature of the Majorana neutrino, lepton number conservation does not constrain the flavor of the light neutrino. Since the light neutrino escapes detection and contributes only to missing transverse energy, summing over its flavor is experimentally motivated. This also enables the exploration of potential lepton-universality violations in these scenarios.

To improve sensitivity, we employ gradient-boosted decision trees (BDTs)~\cite{Friedman:2001wbq}, trained and tested on kinematic observables of simulated events. Using the BDT output, we compute the signal significance and derive the corresponding sensitivity to the heavy--light neutrino mixing parameters, assuming an integrated luminosity of $3000~\mathrm{fb^{-1}}$ expected for the high-luminosity LHC (HL-LHC). Our results indicate that for heavy neutrino masses in the range $50$--$10000~\mathrm{GeV}$, the expected sensitivity to the mixing parameter extends from $\order{10^{-5}}$ to $\order{1}$, representing an improvement over current experimental limits.

The remainder of this paper is organized as follows. In Section~\ref{sec:theory}, we briefly review the seesaw mechanism and define the relevant model parameters. In Section~\ref{sec:simulation}, we describe the generation of signal samples for various heavy neutrino masses, together with the background samples, and present the associated cross sections. In Section~\ref{sec:analysis}, we explain the application of BDTs to event classification and discuss the resulting performance. In Section~\ref{sec:results}, we present the expected sensitivity contours and uncertainty bands for the mixing parameters derived from the BDT analysis under different scenarios. Finally, we conclude by emphasizing that this strategy provides a promising and complementary approach for probing heavy neutrinos in the mass range $50$--$10000~\mathrm{GeV}$.

\section{Theoretical Setup}
\label{sec:theory}

Following the spirit of the seesaw mechanism~\cite{Minkowski:1977sc,Gell-Mann:1979vob,Yanagida:1979as} and the principle of gauge invariance, the simplest renormalizable Lagrangian terms involving neutrino masses are:
\begin{equation}
\mathcal{L}_{\nu} \ni \frac{1}{2}i\bar R_{i}\slashed{\partial}R_{i}-y_{ij}\bar R_{i} \tilde \Phi^\dagger L_{j}-\frac{1}{2}(M_{N})_{ij} \bar R_{i}R^C_{j} + \text{h.c.},
\end{equation}
where the first term represents the kinetic term for the right-handed Majorana neutrino field, $\tilde\Phi$ is the conjugate Higgs doublet, $L_j$ is the $SU(2)_L$ left-handed lepton doublet, $y$ is the leptonic Yukawa coupling matrix, $M_N$ is the Majorana mass matrix, $C$ denotes charge conjugation, and $R_i$ are gauge singlet right-handed neutrino fields. The number of heavy right-handed neutrinos depends on whether the lightest neutrino is massless or not~\cite{Wyler:1982dd,King:2015aea}.

After spontaneous electroweak symmetry breaking, the Dirac mass matrix $(M_D)_{ij}$ for neutrinos arises from the Yukawa interactions. The neutrino mass terms can then be expressed as
\begin{equation}
\frac{1}{2}
\begin{bmatrix}
0 & M^\mathrm{T}_{D}\\
M_{D} & M_{N}
\end{bmatrix}
+ \text{h.c.},
\end{equation}
in the basis of the neutrino flavor eigenstates $\{\nu_L, R^C\}$. Diagonalizing this mass matrix yields three light mass eigenstates ($\nu_i$) and $n$ heavy mass eigenstates ($N_j$). The corresponding weak interaction terms are given by
\begin{equation}\label{eq:l}
\begin{split}
\mathcal{L} \ni &-\frac{g}{\sqrt{2}}(U^{*}_{\ell i}\bar \nu_{i}+ V^{*}_{\ell j}\bar N_{j}) \slashed{W}^{+} P_L\ell + \text{h.c.} \\
&-\frac{g}{2\cos\theta _{W}}(U^{*}_{\ell i}\bar \nu_{i}+ V^{*}_{\ell j}\bar N_{j}) \slashed{Z} P_L(U_{\ell m}\nu_{m}+V_{\ell n}N_{n}),
\end{split}
\end{equation}
where $g$ is the gauge coupling constant of the $SU(2)_L$ group, $U_{\ell i}$ ($V_{\ell j}$) are the mixing parameters between the weak interaction eigenstates and the light (heavy) mass eigenstates, $\theta_W$ is the weak-mixing angle, and $P_L$ is the left-handed projection operator. The mixing parameters $V_{\ell j}$ can be small~\cite{Minkowski:1977sc,Schechter:1980gr}, such that the matrix $U_{\ell i}$ is approximately unitary, {\it i.e.}, $U^\dagger U \approx \mathbf{1}_{3 \times 3}$, and can be identified with the Pontecorvo--Maki--Nakagawa--Sakata (PMNS) matrix~\cite{Pontecorvo:1957qd,Maki:1962mu,Pontecorvo:1967fh,ParticleDataGroup:2024cfk}. Since our projected sensitivity can reach $|V_{\ell j}| \sim 1$ in specific mass ranges, we perform a comprehensive analysis of the full $|V_{\ell j}|$ phase space, extending up to unity.

We note that scenarios with heavy Majorana neutrinos can also provide viable explanations for the persistent $R(D^{(*)})$ anomalies in semileptonic $B$-meson decays~\cite{BaBar:2012obs,BaBar:2013mob,LHCb:2015gmp}. In particular, scenarios in which the heavy Majorana neutrino couples predominantly to the $\tau$ flavor--while its mixings with the electron and muon sectors remain suppressed--can naturally induce the required modifications to the charged-current transition amplitudes~\cite{Atre:2009rg,Calibbi:2015kma}. This $\tau$-enhanced structure is not only theoretically well-motivated but also phenomenologically attractive, as current collider constraints are significantly weaker for models dominated by $\tau$ couplings~\cite{DELPHI:1996qcc,CMS:2024xdq,ATLAS:2024fcs,CMS:2016fxb,CMS:2018iye}.

Motivated by these considerations, we first examine a benchmark model in which the heavy Majorana neutrino couples exclusively to a single lepton flavor. We then extend our analysis to a more general framework in which $N$ couples to all three charged-lepton generations, thereby allowing for potential lepton-universality violation~\cite{Cai:2017mow}.

\section{Sample Simulations}
\label{sec:simulation}

The signal and background events are simulated and analyzed using \verb|FeynRules|~\cite{Alloul:2013bka} with the HeavyN model files~\cite{Alva:2014gxa,Degrande:2016aje}, \verb|MadGraph5_aMC@NLO 3.5.6|~\cite{Alwall:2014hca,Frederix:2018nkq}, and \verb|MadWidth|~\cite{Alwall:2014bza}. The NNPDF2.3 LO parton distribution function set~\cite{Ball:2012cx} is employed for event generation, with the proton--proton collision energy fixed at $\sqrt{s} = 13.6~\mathrm{TeV}$.

At the parton level, all samples are generated using identical selection requirements. Charged leptons are required to have transverse momentum $p_\mathrm{T} > 10~\mathrm{GeV}$, while jets must satisfy $p_\mathrm{T} > 20~\mathrm{GeV}$, $|\eta| < 5$. The pseudorapidity of charged leptons is restricted to $\abs{\eta} < 2.5$, and for jets to $\abs{\eta} < 5$. The cross sections of all samples are taken directly from the \verb|MadGraph5_aMC@NLO| output, and calculated at LO in QCD.

\begin{figure}[htbp]
    \centering
    \begin{minipage}[b]{0.45\linewidth}
        \centering
        \includegraphics[width=\linewidth,keepaspectratio]{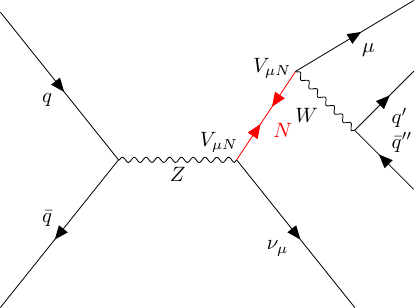}
    \end{minipage}
    \hspace{0.1\linewidth}
    \begin{minipage}[b]{0.35\linewidth}
        \centering
        \includegraphics[width=\linewidth,keepaspectratio]{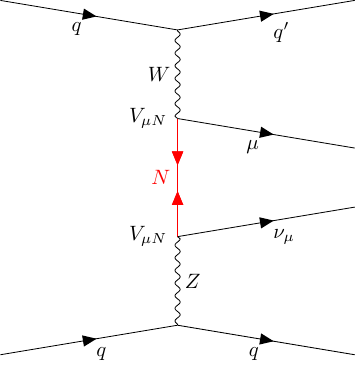}
    \end{minipage}
    \caption{Representative Feynman diagrams illustrating the production of
    $\mu \nu_\mu jj$ via $s$ channel (left) and VBF processes (right),
    mediated by a heavy neutrino $N$. The parameter $V_{\mu N}$ denotes
    the corresponding mixing parameter.}
    \label{fig: Feyn}
\end{figure}

\begin{figure}[htbp]
\begin{center} 
  \includegraphics[width=0.45\textwidth, keepaspectratio]{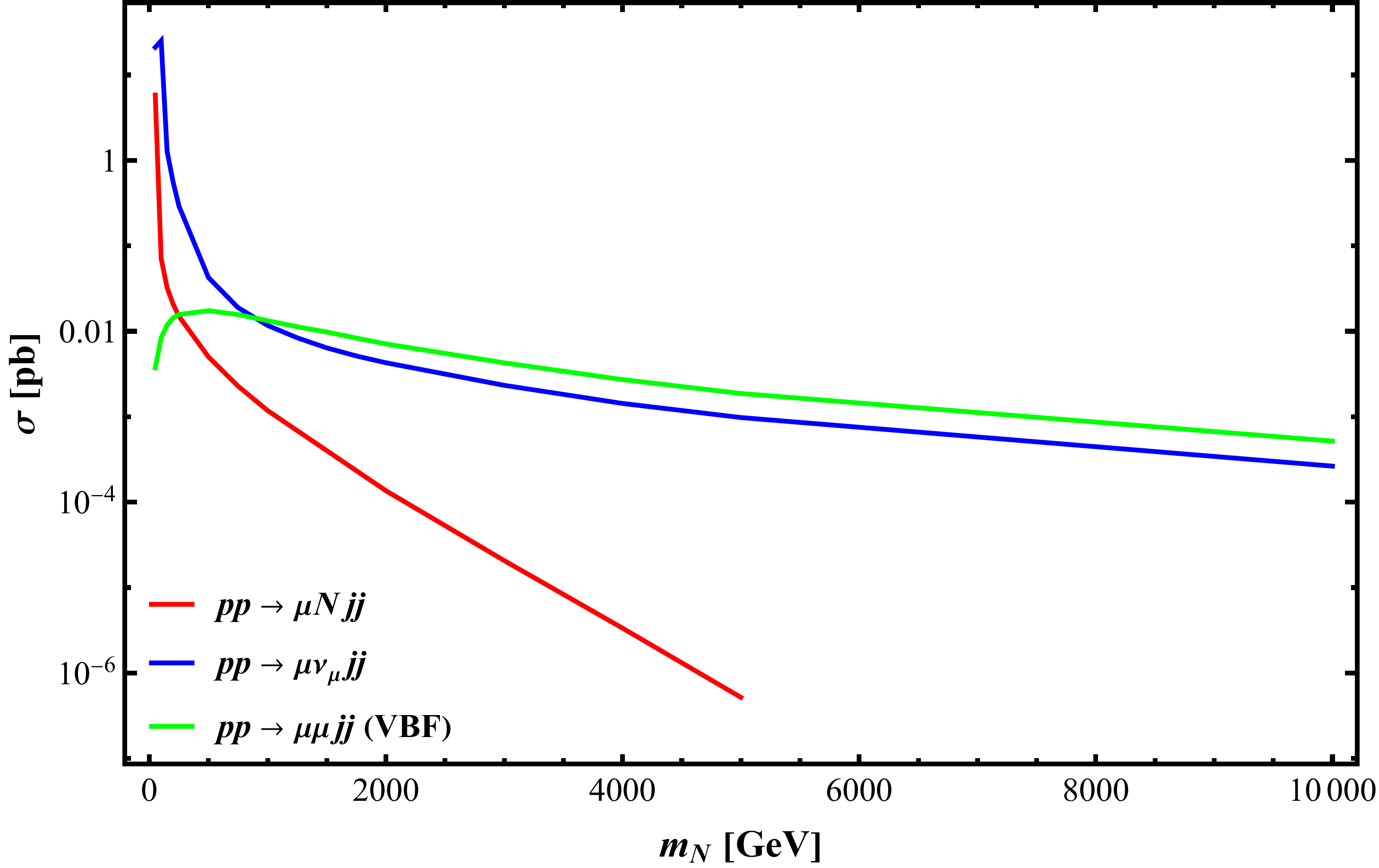}
\end{center}
\caption{Production cross sections (with $V_{\ell N} = 1$) as functions of the heavy neutrino mass $m_N$ at a center-of-mass energy of $\sqrt{s} = 13.6~\mathrm{TeV}$. The red curve shows the cross section for the process $pp \to \mu N jj$. The blue curve corresponds to the inclusive cross section for $pp \to \mu \nu_\mu jj$ while the green curve denotes the VBF-induced cross section for the same-sign dilepton channel $pp \to \mu \mu jj$, as used in Refs.~\cite{CMS:2022hvh,ATLAS:2024rzi}. The three curves converge near $m_N \simeq 1~\mathrm{TeV}$, beyond which the VBF contribution becomes dominant. Our numerical results are consistent with Ref.~\cite{Fuks:2020att}.}
\label{fig: Cross Section}
\end{figure}

Signal samples are generated via pure electroweak processes $pp \to \ell \nu jj$ at order $\order{g^4}$. Taking $\ell = \mu$ as a representative case, we illustrate the relevant Feynman diagrams in FIG.~\ref{fig: Feyn} and present the signal cross section for the process $pp \to \mu \nu_\mu jj$ as a function of $m_N$ in FIG.~\ref{fig: Cross Section}. As illustrated in Fig.~\ref{fig: Cross Section}, for lower heavy neutrino masses, the heavy neutrino can be resonantly produced through $s$ channel processes such as Drell--Yan and $W\gamma$ fusion~\cite{Alva:2014gxa,Florez:2017xhf}. Compared to searches for heavy neutrinos via VBF processes leading to same-sign dilepton final states~\cite{CMS:2022hvh,ATLAS:2024rzi}, $s$ channel production yields larger cross sections at low masses. However, these contributions become kinematically suppressed as $m_N$ increases. In contrast, VBF-like topologies involving virtual heavy neutrino exchange (as shown in the right panel of Fig.~\ref{fig: Feyn}) exhibit a milder dependence on the heavy neutrino mass. Within the inclusive $\ell\nu jj$ final state considered in this work, these production topologies dominate in complementary mass regimes, potentially enabling sensitivity to heavy neutrinos over a broad mass range at the LHC and to smaller active--sterile mixing parameters.

Signal samples are generated for 16 heavy neutrino mass points, ranging from 50 GeV to 10 TeV, with varying step sizes. For the SM background, we consider several sources, including single top and top-quark pair production, single Higgs production (via gluon-gluon fusion, VBF, and associated production), vector boson production $V=\{Z, W\}$ in association with jets, pure QCD multijet events, diboson production ($VV=\{WW, ZZ, WZ\}$), and triboson processes (${VVV}$). Among these, the dominant backgrounds arise from $V$ + jets and diboson production with associated jets. The dominant SM background is generated as an inclusive $pp\to \ell\nu+\geq 2j$ sample using the default SM model. No additional coupling-order restriction is imposed by hand. With the default MadGraph coupling-order selection, this sample is dominated by QCD-induced $W+$jets production, with two electroweak vertices associated with the $W\to \ell\nu$ decay and multiple QCD vertices associated with the two or more jets. The resulting inclusive background cross section is $1888~\mathrm{pb}$. All remaining background processes listed above are found to be subdominant and do not contribute appreciably to our analysis.

Although not included among the dominant simulated background processes, we have also considered the potential impact of QCD multijet production. This background corresponds primarily to multijet events with at least three jets $(p p \rightarrow \geq 3j)$, where one jet is misidentified as an isolated charged lepton and jet energy mismeasurement gives rise to fake missing transverse momentum. While the inclusive QCD multijet production cross section is several orders of magnitude larger than that of the dominant $pp\to \ell\nu+\geq 2j$ background, the baseline event selection (discussed later) suppresses the QCD multijet contribution by several orders of magnitude, making it a small contribution with respect to the other dominant backgrounds. Thus, the QCD multijet background is not included as a dominant process in these studies. 

The generated parton-level events are subsequently passed to $\verb|Pythia 8.2.30|$~\cite{Sjostrand:2014zea}, where parton shower evolution and hadronization are modeled. Detector-level effects are then emulated with $\verb|Delphes 3.4.1|$~\cite{deFavereau:2013fsa}, employing CMS detector configurations for object reconstruction and particle identification. The charged-hadron, electron, and muon reconstruction efficiencies, as well as the corresponding misidentification rates and the b-tagging efficiencies and mistag rates quoted below, are implemented directly within the Delphes detector card and are therefore incorporated as part of the detector simulation. Jets are reconstructed using the anti-$k_\mathrm{T}$ clustering scheme~\cite{Cacciari:2008gp} with a radius parameter of $R=0.4$ in $\verb|FastJet 3.4.2|$~\cite{Cacciari:2011ma}. Jet matching and merging are performed using the MLM prescription~\cite{Mangano:2006rw}, with the matching scales chosen as \textsc{xqcut} = 30 and \textsc{qcut} = 45 to maintain a smooth behavior of the differential jet rates across different jet multiplicities.

Efficient reconstruction of leptons, jets initiated by light quarks or gluons, and bottom-flavored jets plays a central role in isolating signal processes from SM backgrounds and is therefore essential for maximizing sensitivity at the HL-LHC. At the same time, the large number of overlapping proton--proton interactions anticipated at the HL-LHC presents significant challenges for particle and object identification. The impact of pileup on searches for physics beyond the SM, as well as the effectiveness of pileup mitigation strategies for the CMS and ATLAS experiments, has been studied in Ref.~\cite{CMS:2013ega}. Although a detailed evaluation of the upgraded detector performance lies beyond the scope of this analysis, we adopt conservative performance assumptions by allowing for a modest reduction in lepton and hadron identification efficiencies. These assumptions are guided by Ref.~\cite{CMS:2013ega} and correspond to an average of 140 pileup interactions per bunch crossing.

The efficiency for reconstructing charged hadrons, which directly impacts both jet/tau reconstruction and the determination of missing transverse momentum, is taken to be approximately 97\% in the pseudorapidity range $|\eta| < 1.5$, decreasing to about 85\% at $|\eta| = 2.5$. For electrons and muons with transverse momentum $p_\mathrm{T} > 5$ GeV and $|\eta| < 1.5$, a baseline identification efficiency of 95\% is assumed, together with a misidentification probability of 0.3\%~\cite{CMS:2013ega,CMS:2019ied}. In the forward region, $1.5 < |\eta| < 2.5$, the performance is assumed to degrade linearly, reaching an identification efficiency of 65\% and a misidentification rate of 0.5\% at the edge of the tracker acceptance. These performance losses, primarily driven by additional proton--proton interactions, result in reduced precision in the reconstructed lepton kinematics. For bottom-quark identification, we follow Ref.~\cite{CMS:2017wtu} and modify the $\verb|Delphes 3.4.1|$ card to employ the ``Loose" working point of the DeepCSV tagger~\cite{Bols:2020bkb}, which corresponds to a $p_{\textrm{T}}$-dependent $b$-tagging efficiency of approximately 70--85\% and a light-flavor mistag rate of about 10\%. The specific choice of the $b$-tagging operating point is determined through an optimization procedure aimed at suppressing top-quark backgrounds and maximizing the overall discovery sensitivity.

A comprehensive study to address suitable and available HL-LHC triggers is beyond the scope of this work. However, we note that the CMS compressed SUSY search in~\cite{CMS:2018kag} utilized a soft lepton plus $E^{\mathrm{miss}}_{\textrm{T}}$ trigger. As such, we deem it reasonable that an HL-LHC trigger be developed for the proposed search strategy in this paper.

\section{Data Analysis}
\label{sec:analysis}

The analysis of signal and background events is performed using a machine-learning--based classifier known as gradient-boosted decision trees (BDTs)~\cite{Friedman:2001wbq}, which have become a standard tool in high-energy physics for multivariate classification tasks. BDTs are ensemble learning methods that combine the outputs of multiple decision trees trained sequentially to enhance predictive performance. Each tree in the sequence is trained to correct the errors made by its predecessors, with the final model constructed as a weighted sum of the individual trees. This gradient-boosting approach effectively minimizes a specified loss function in a greedy, iterative manner, allowing the model to capture complex nonlinear relationships in the input data.

One of the main strengths of BDTs lies in their ability to incorporate a large number of input features simultaneously and to automatically identify the most informative combinations. In the context of particle physics, this capability is particularly advantageous for event-classification problems, where signal and background processes may differ subtly across multiple kinematic variables. By exploiting correlations and higher-order dependencies among these variables, BDTs enable the construction of highly discriminative decision boundaries that go beyond traditional cut-based analyses. Their interpretability, flexibility, and robustness have made them a widely adopted choice in collider experiments such as those at the LHC~\cite{Roe:2004na,CMS:2013poe,Baldi:2014kfa,ATLAS:2017fak,Albertsson:2018maf,Barbosa:2022mmw,Qureshi:2024naw}, where distinguishing rare signals from overwhelming backgrounds is a central challenge.

For our analysis, we employed the BDT implementations provided by the \verb|Scikit-learn| and \verb|XGBoost| libraries~\cite{Pedregosa:2011ork,Chen:2016btl}. Specifically, we used the \verb|XGBClassifier| with the objective function set to \verb|binary: logistic|, which models the probability of a binary outcome using logistic regression. Our analysis pipeline relies on a dedicated \verb|MadAnalysis| expert-mode C++ framework designed to process the simulated event samples and extract the relevant kinematic and event-level observables. The selected variables are reformatted and exported into a structured CSV dataset, which serves as the input for the subsequent machine-learning training. To correctly represent the relative contributions of signal and background processes, event weights derived from their respective production cross sections are applied, ensuring a statistically consistent dataset.

The weighted signal and background samples are then reprocessed within the \verb|MadAnalysis| expert-mode environment, where an initial event selection is performed prior to CSV generation. Events are categorized into independent analysis channels according to the reconstructed lepton flavor, with each flavor treated as a separate search channel and associated with its own independently trained and validated BDT classifier. Reconstructed object-level selection requirements are imposed as follows: reconstructed electrons and muons (hadronically decaying taus) are required to satisfy $p_\mathrm{T} > 10\,(15)~\mathrm{GeV}$ and $|\eta| < 2.5$. To suppress backgrounds containing heavy-flavor production, events with at least one identified b-tagged jet with $p_\mathrm{T} > 30~\mathrm{GeV}$ and $|\eta| < 2.5$ are vetoed. The target signal topology includes two additional hadronic jets and genuine missing momentum; consequently, a dijet selection is enforced by requiring a minimum of two jets with transverse momentum $p_\mathrm{T} > 30~\mathrm{GeV}$ and pseudorapidity $|\eta| < 5$, and a minimum missing transverse momentum of $20$ GeV. All jet pairs satisfying these criteria are subsequently combined to construct dijet candidates, and the baseline selection requires at least one dijet candidate with $m_{jj} > 50$ GeV. This baseline event selection requiring one isolated lepton, multiple jets in the relevant kinematic region, at least one good dijet candidate, and a minimum missing transverse momentum suppresses the QCD multijet contribution by approximately $1.5 \times 10^{-6}$. After these selections, its expected yield is only about 5\% of the dominant $pp\to \ell\nu+\geq 2j$ background. The residual QCD multijet background surviving the baseline selection populates predominantly the low-output region of the BDT discriminant, where the events are readily separated from the signal. Consequently, the signal-enriched region used to derive the projected sensitivities is dominated by electroweak backgrounds, primarily 
$pp\to \ell\nu+\geq 2j$, while the QCD multijet contribution becomes negligible.

Each model was trained with 31 kinematic features: the transverse momentum ($p_\mathrm{T}$), longitudinal momentum ($p_z$), azimuthal angle ($\phi$), and pseudorapidity ($\eta$) of the two jets and the charged lepton; the missing transverse energy ($E^{\mathrm{miss}}_\mathrm{T}$); the azimuthal angle of the transverse missing energy; the azimuthal angle separation ($\Delta \phi$), pseudorapidity separation ($\Delta \eta$), and angular separation ($\Delta R$) between the two jets and between the missing transverse energy and the charged lepton; the pseudorapidity separation between the leading jet and the charged lepton; the transverse mass $M_\mathrm{T}$ of each visible object and missing transverse energy; visible transverse energy $E_\mathrm{T}$; hadronic transverse energy $H_\mathrm{T}$; the invariant mass ($M$) of the dijet system and charged lepton; and the relative transverse momentum between the leading jet and the charged lepton, defined as {$(p_\mathrm{T}[j_1] - p_\mathrm{T}[\ell])/p_\mathrm{T}[j_1]$}. The two jets are ranked by their transverse momentum, such that $p_\mathrm{T}[j_1]$ and $p_\mathrm{T}[j_2]$ represent the leading and subleading jets, respectively.

A separate model was trained for each heavy neutrino mass, using one million signal and background events per mass point. An additional 0.5 million events were used for hyperparameter tuning via Optuna~\cite{Akiba:2019lwq} with 200 trials, and a further 0.5 million events are reserved for final model validation.

\section{Results}
\label{sec:results}

\begin{figure*}[t]
    \centering
    \includegraphics[width=0.32\textwidth]{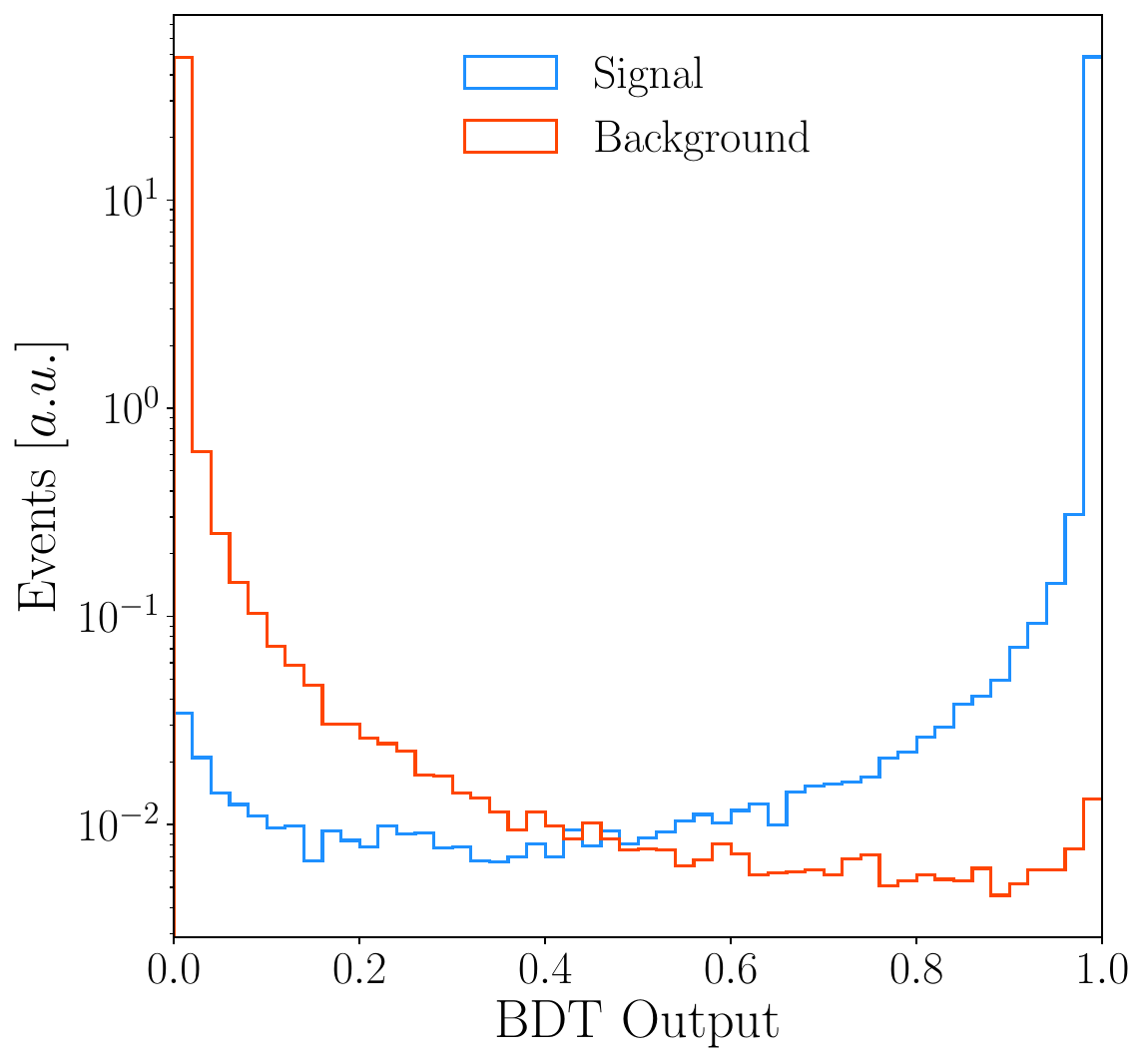}\hfill
    \includegraphics[width=0.32\textwidth]{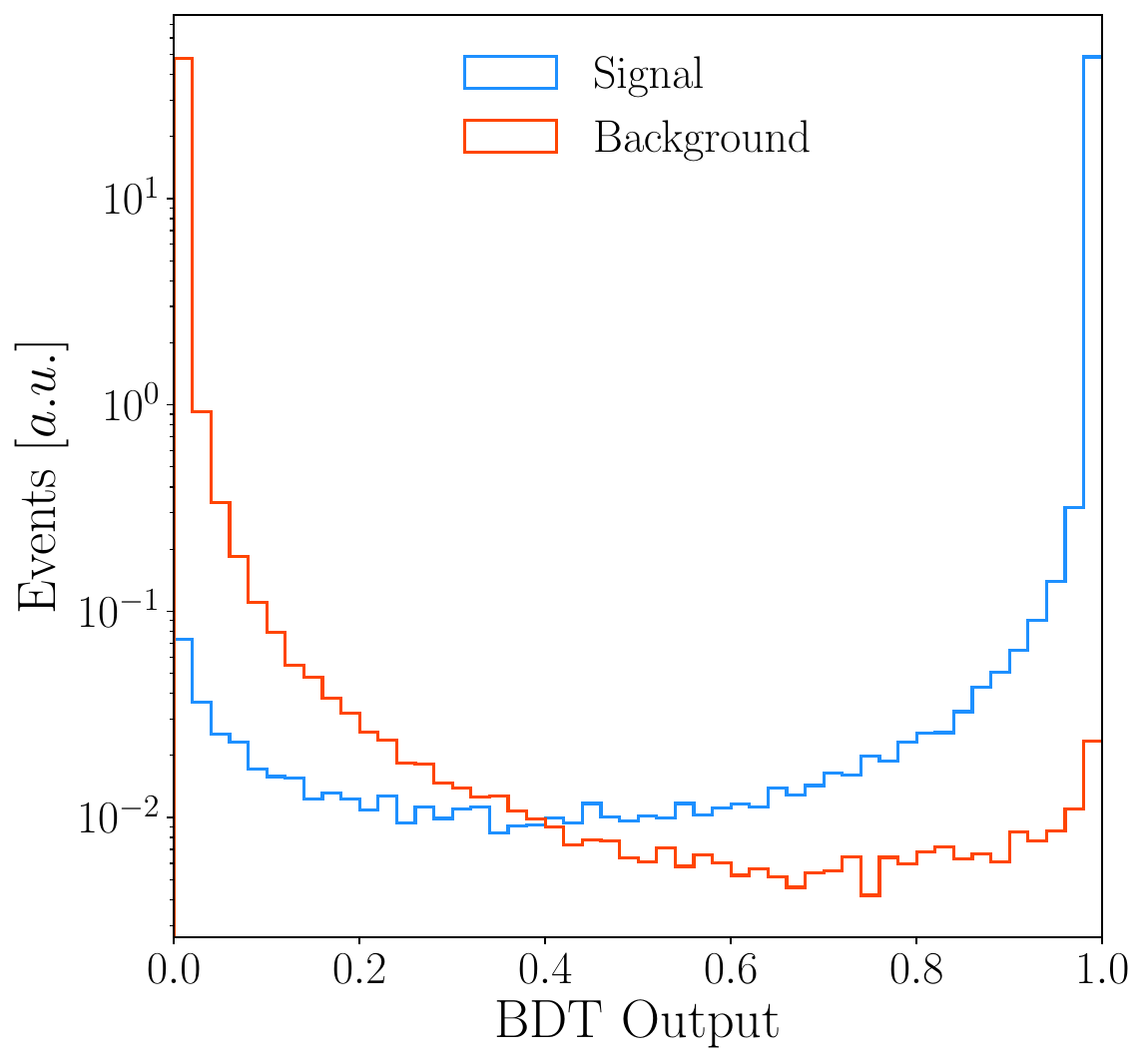}\hfill
    \includegraphics[width=0.32\textwidth]{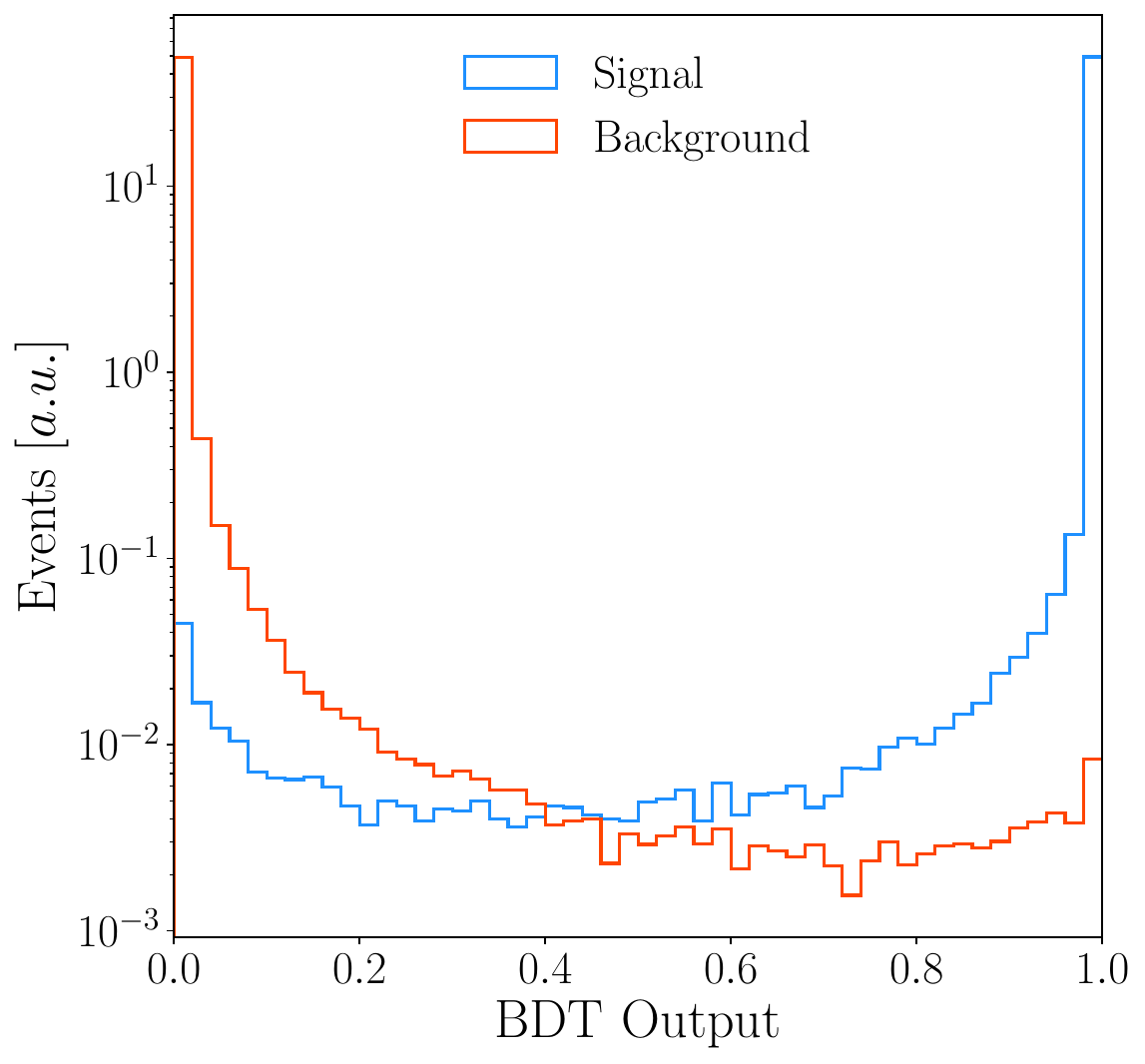}
    \caption{BDT output distributions (normalized to 1) for signal and background events at heavy neutrino masses of 0.1 TeV (left), 1 TeV (middle), and 10 TeV (right) for $\ell = \mu$. The vertical axis shows the normalized event yield in arbitrary units, and the horizontal axis represents the BDT output score.}
    \label{fig: BDT}
\end{figure*}

FIG.~\ref{fig: BDT} presents the BDT output distributions for heavy neutrino masses of 0.1, 1, and 10 TeV in the process $pp \to \mu \nu_\mu jj$. During BDT training, signal and background events are labeled as 1 and 0, respectively. Therefore, a score closer to 0 corresponds to background-like events, while a score closer to 1 corresponds to signal-like events.

\begin{figure*}[t]
    \centering
    \includegraphics[width=0.32\textwidth]{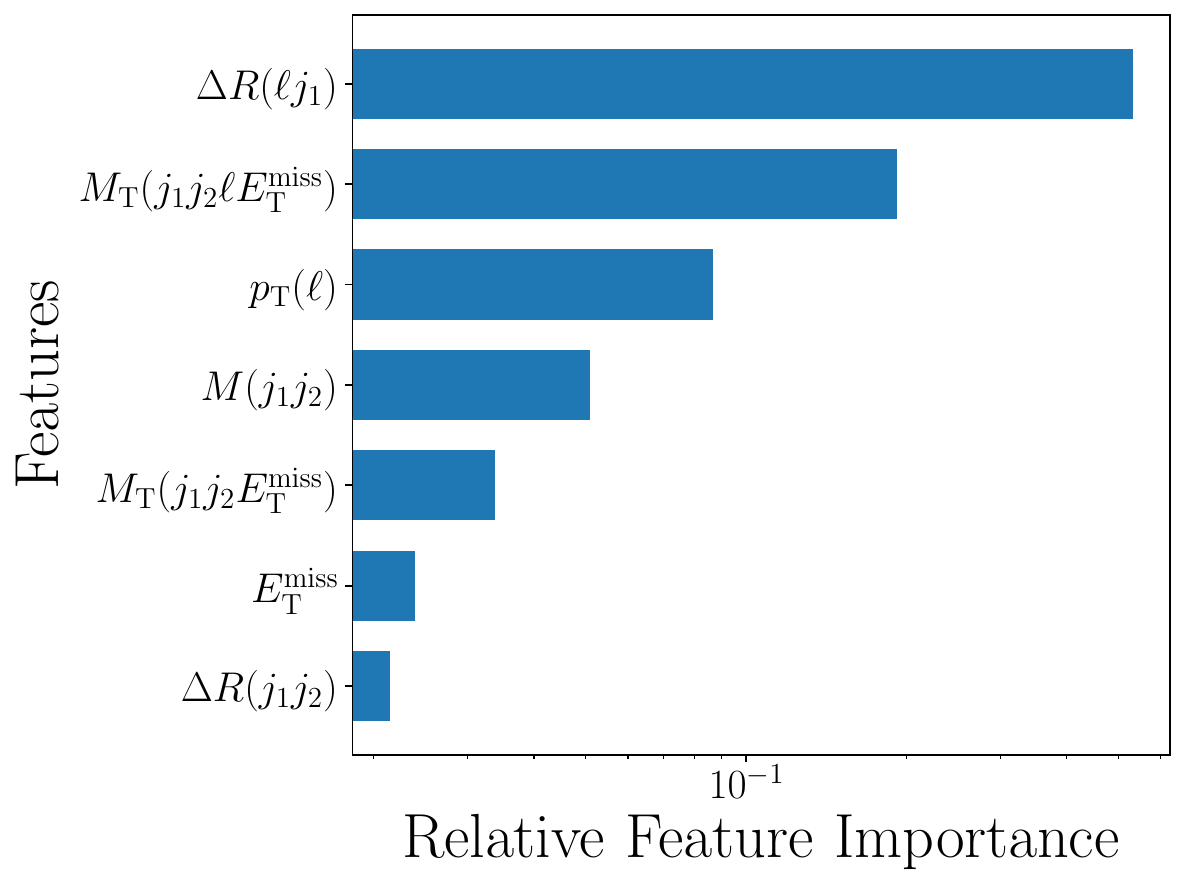}\hfill
    \includegraphics[width=0.32\textwidth]{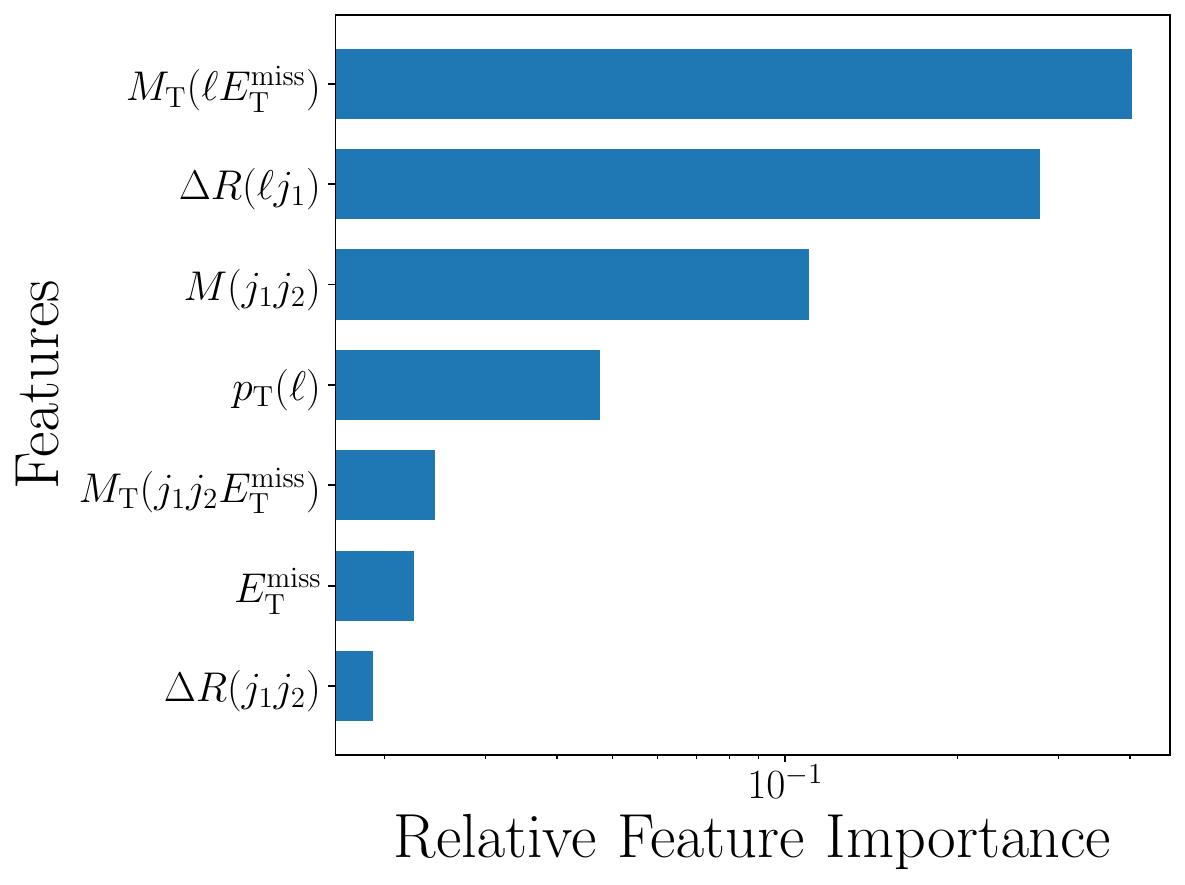}\hfill
    \includegraphics[width=0.32\textwidth]{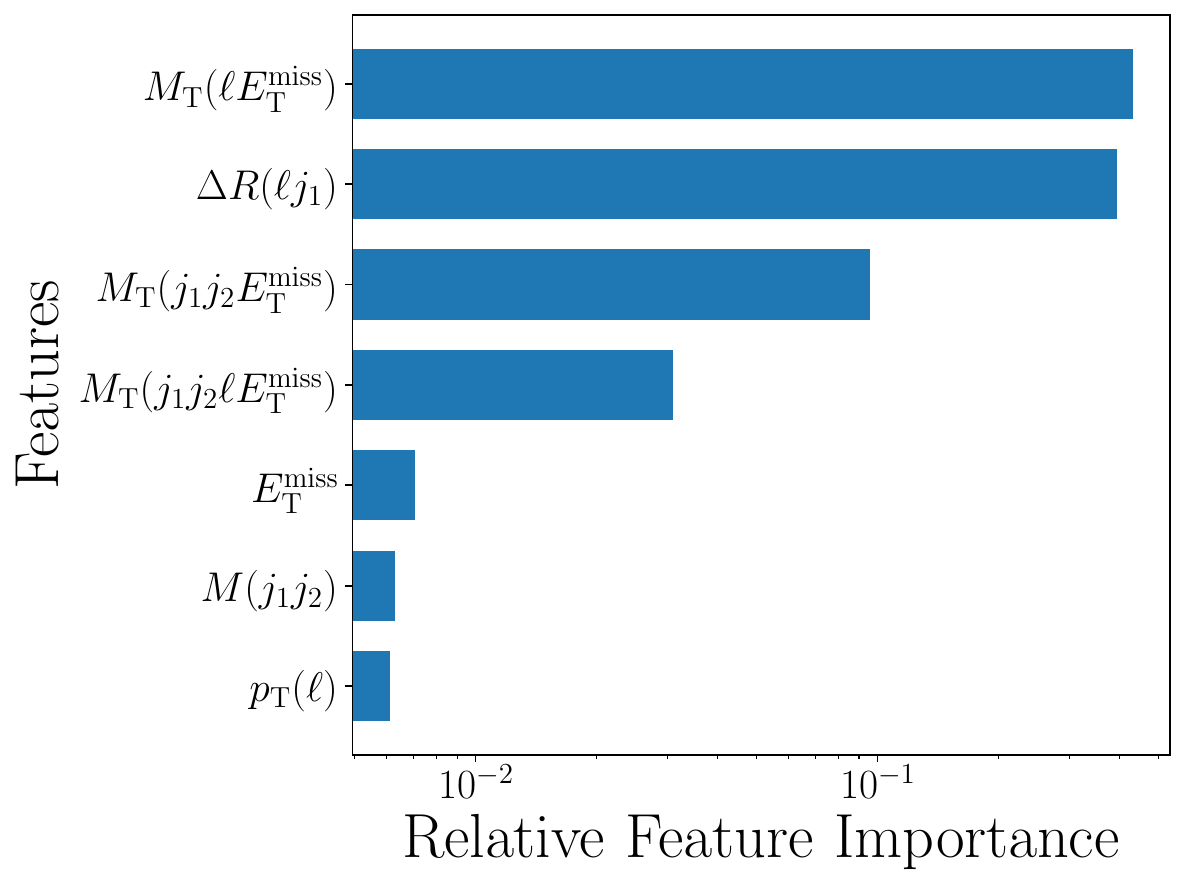}
    \caption{Relative feature importance for the BDT classifier at a heavy neutrino mass of 0.1 TeV (left), 1 TeV (middle), and 10 TeV (right) for $\ell = \mu$.}
    \label{fig: BDT_Relative}
\end{figure*}

\begin{figure}[htbp]
    \begin{center} 
        \includegraphics[width=0.45\textwidth, keepaspectratio]{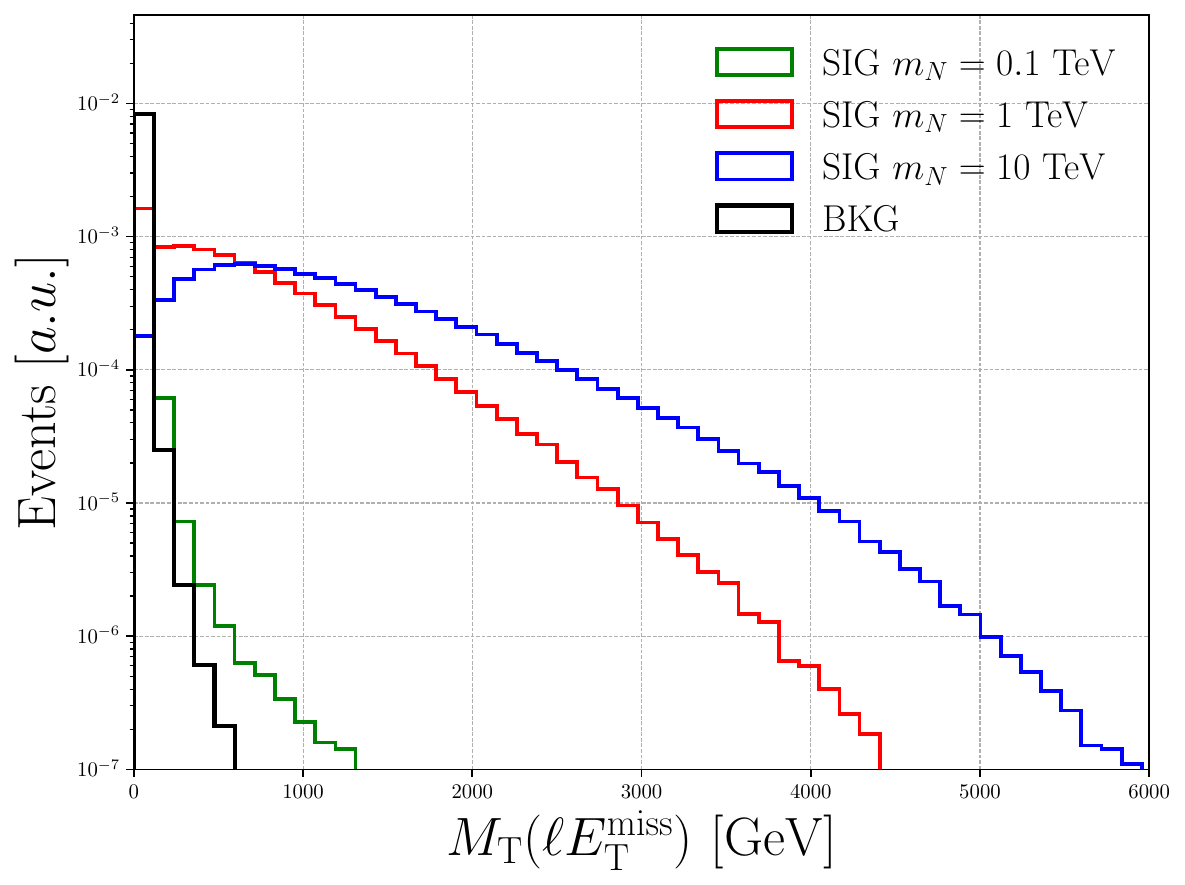}
    \end{center}
\caption{Distribution (normalized to 1) of the variable $M_\mathrm{T}(\ell E^\mathrm{miss}_\mathrm{T})$ for background (BKG) and signal (SIG) events corresponding to heavy neutrino masses of 0.1, 1, and 10 TeV, shown together in a single panel for $\ell = \mu$.}
\label{fig: MML}
\end{figure}

The corresponding relative feature importance is shown in FIG.~\ref{fig: BDT_Relative}. The importance reflects each variable’s contribution to the model’s decision-making process, with kinematic features ranked according to their discriminating power between signal and background events. As shown in FIG.~\ref{fig: BDT_Relative}, one of the most important features is the transverse mass constructed from the missing energy and the charged lepton. For SM background processes, these two objects are most likely produced from the same $W$ boson, resulting in a transverse mass distribution that peaks near the $W$ boson mass. This behavior is confirmed by the corresponding event distribution shown in FIG.~\ref{fig: MML}.

\begin{figure}[htbp]
    \begin{center} 
  \includegraphics[width=0.45\textwidth, keepaspectratio]{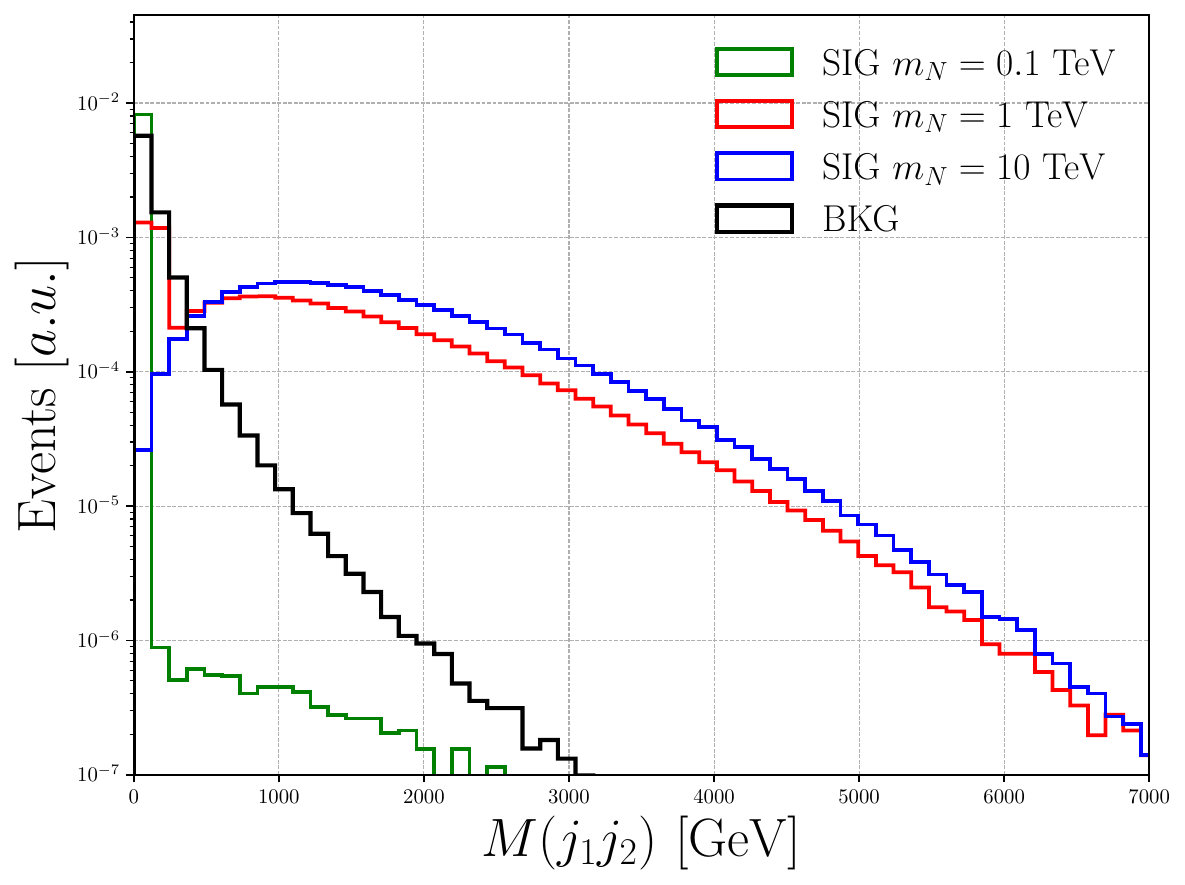}
\end{center}
\caption{Distribution (normalized to 1) of the variable $M(j_1 j_2)$ for background (BKG) and signal (SIG) events corresponding to heavy neutrino masses of 0.1, 1, and 10 TeV, shown together in a single panel for $\ell = \mu$.}
\label{fig: Mjj}
\end{figure}

In VBF processes, two forward/backward jets are typically produced, while the charged lepton and missing energy are located in the central region. As a result, the dijet invariant mass $M(j_1j_2)$ provides discrimination power. As shown in FIG.~\ref{fig: Mjj}, for lower heavy neutrino masses, a sharp peak appears in the range $M(j_1j_2)\sim 100~\mathrm{GeV}$. This feature originates from resonant heavy neutrino production via the $s$ channel, followed by its decay into an on-shell $W$ or $Z$ boson that subsequently decays into a dijet system. For larger heavy neutrino masses, the $t$ channel VBF process becomes dominant, leading to a smoother $M(j_1 j_2)$ distribution with larger invariant masses typical of VBF processes~\cite{Delannoy:2013ata,CMS:2016ucr,CMS:2024vhy,Qureshi:2024cmg,Cardona:2021ebw,Florez:2021zoo,CMS:2019san,Florez:2019tqr,Florez:2018ojp,Florez:2016uob,CMS:2015jsu,Dutta:2015hra,Dutta:2014jda,Dutta:2013gga,Dutta:2012xe}.

\begin{figure}[htbp]
    \begin{center} 
        \includegraphics[width=0.45\textwidth, keepaspectratio]{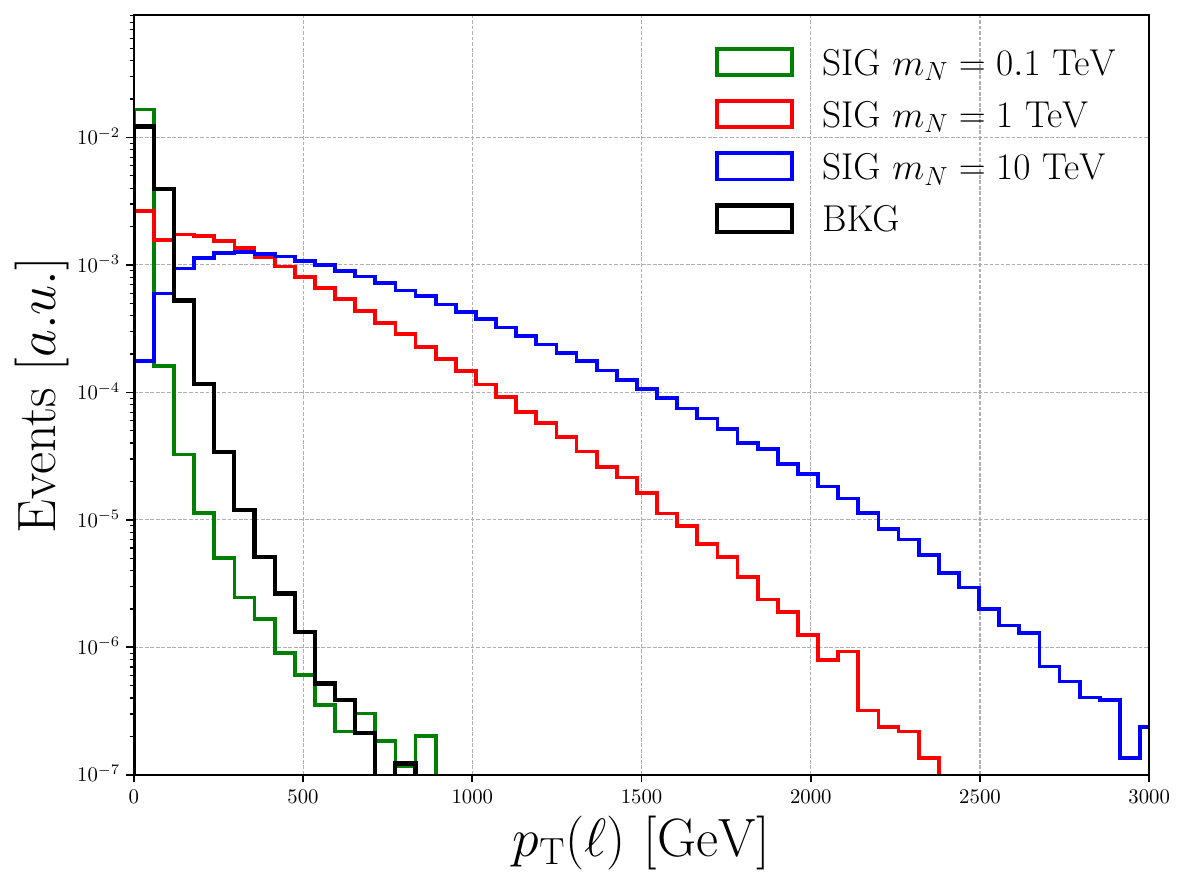}
    \end{center}
\caption{Distribution (normalized to 1) of the variable $p_\mathrm{T}(\ell)$ for background (BKG) and signal (SIG) events corresponding to heavy neutrino masses of 0.1, 1, and 10 TeV, shown together in a single panel for $\ell = \mu$.}
\label{fig: pTL}
\end{figure}

At higher heavy neutrino masses, the VBF jets also exhibit larger angular separation, while the charged lepton and missing energy acquire larger transverse kinematic features such as $p_\mathrm{T}$, $E_\mathrm{T}$, and $M_\mathrm{T}$. These variables provide additional discrimination power against the background and are effectively captured by the BDT model, {\it e.g.}, as illustrated in FIG.~\ref{fig: pTL}.

\begin{figure*}[t]
    \centering
    \includegraphics[width=0.32\textwidth]{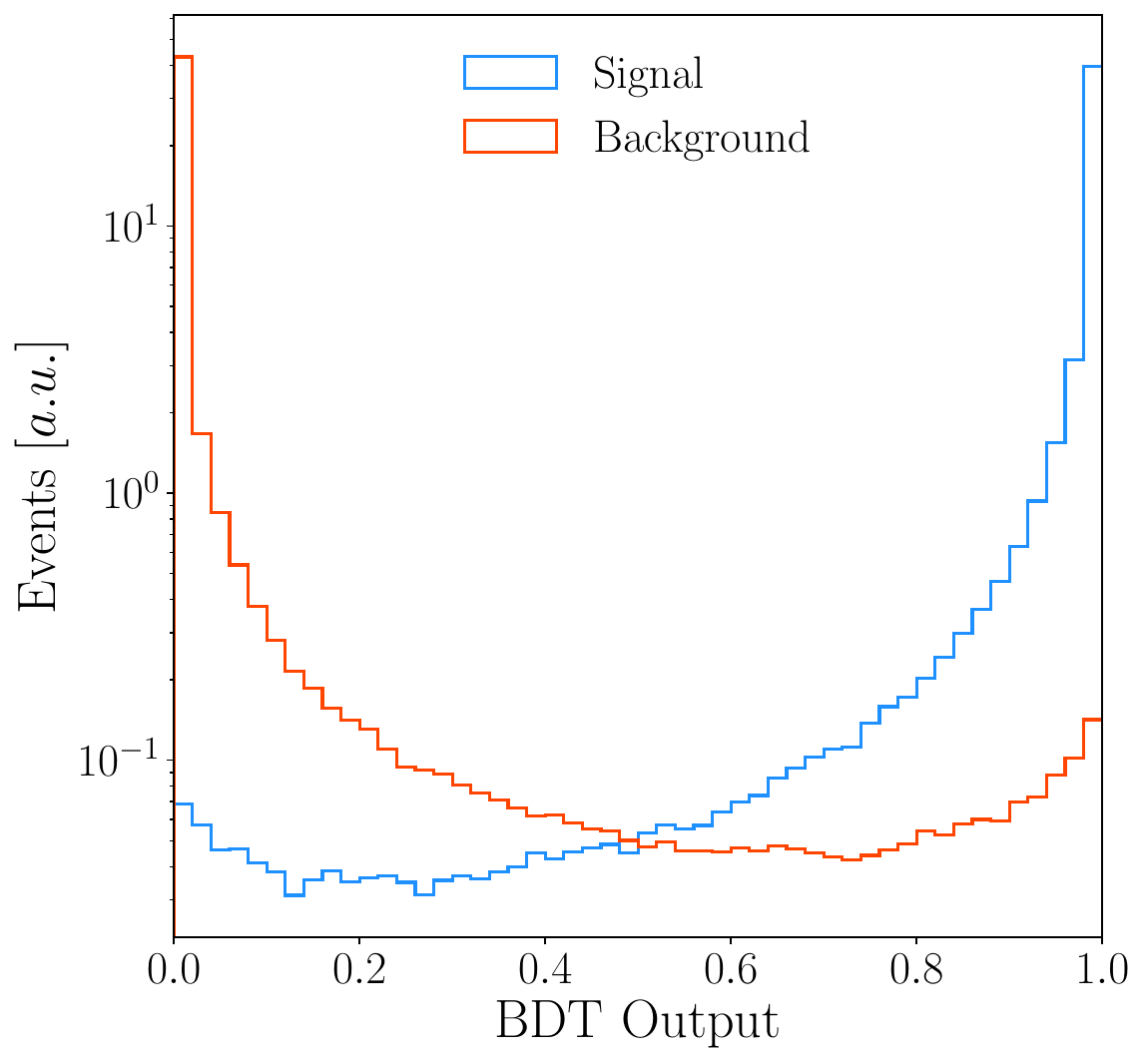}\hfill
    \includegraphics[width=0.32\textwidth]{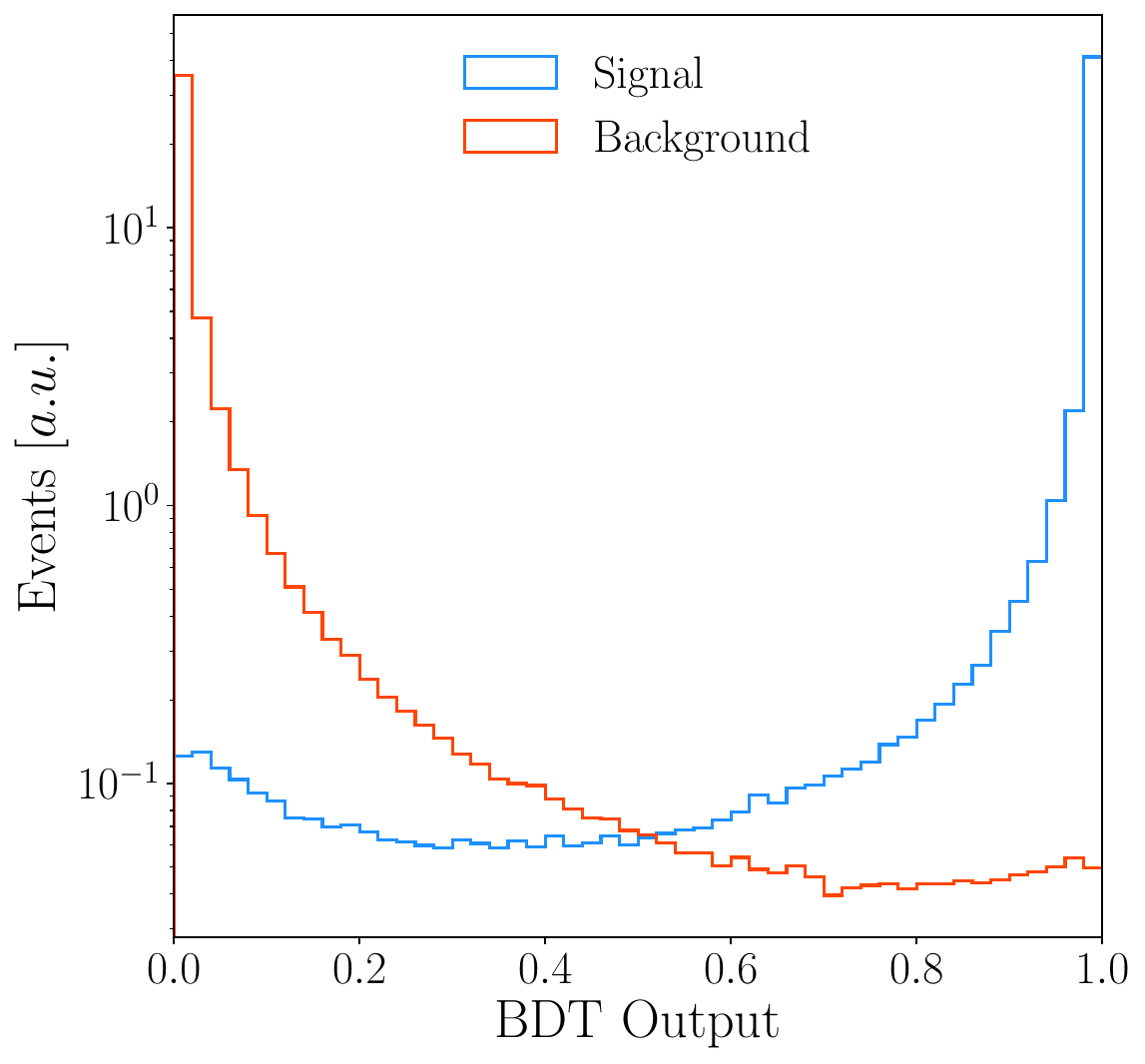}\hfill
    \includegraphics[width=0.32\textwidth]{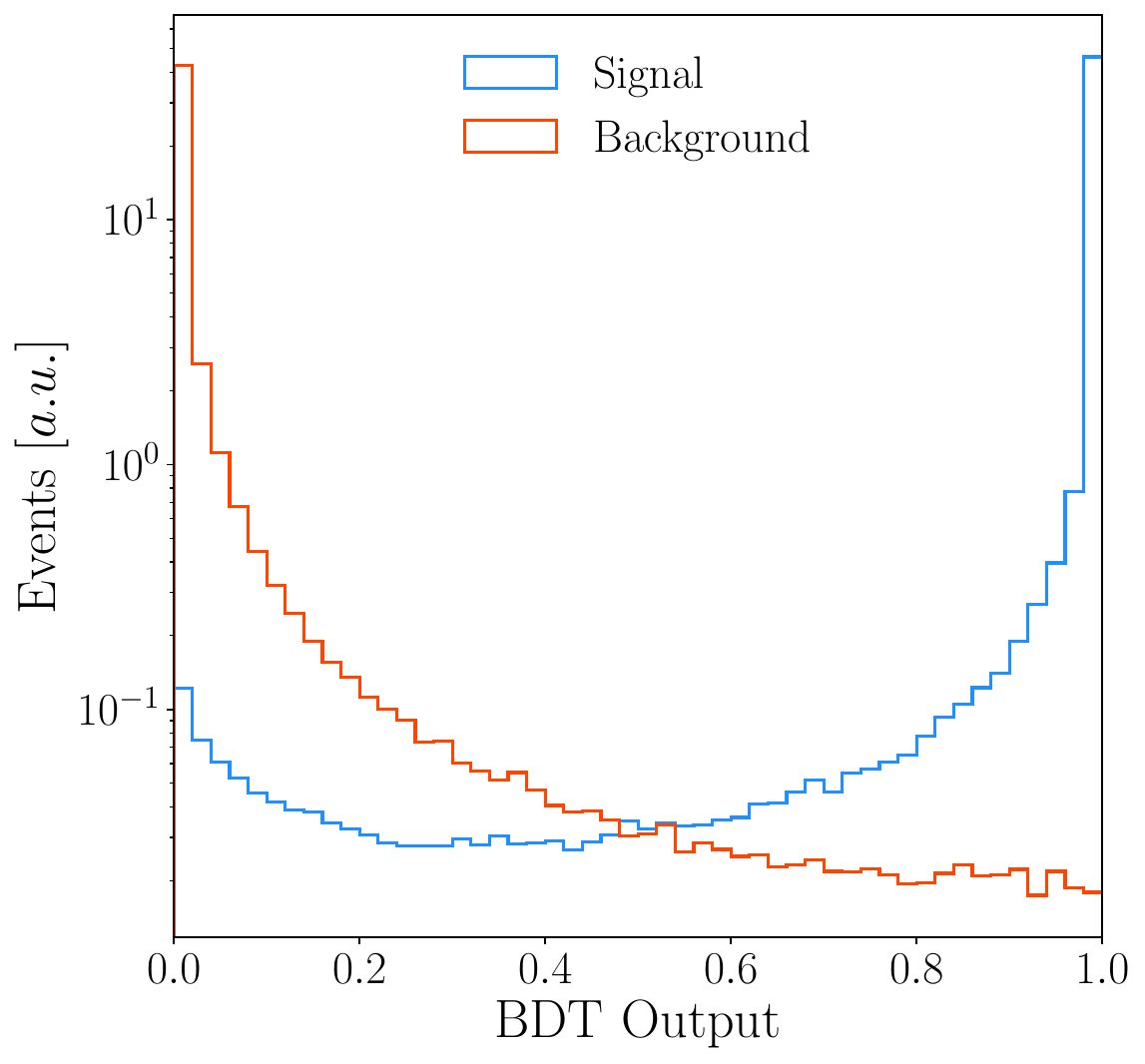}
    \caption{BDT output distributions (normalized to 1) for signal and background events at heavy neutrino masses of 0.1 TeV (left), 1 TeV (middle), and 10 TeV (right) for $\ell = \tau$. The vertical axis shows the normalized event yield in arbitrary units, and the horizontal axis represents the BDT output score.}
    \label{fig: BDT_tau}
\end{figure*}

\begin{figure*}[t]
    \centering
    \includegraphics[width=0.32\textwidth]{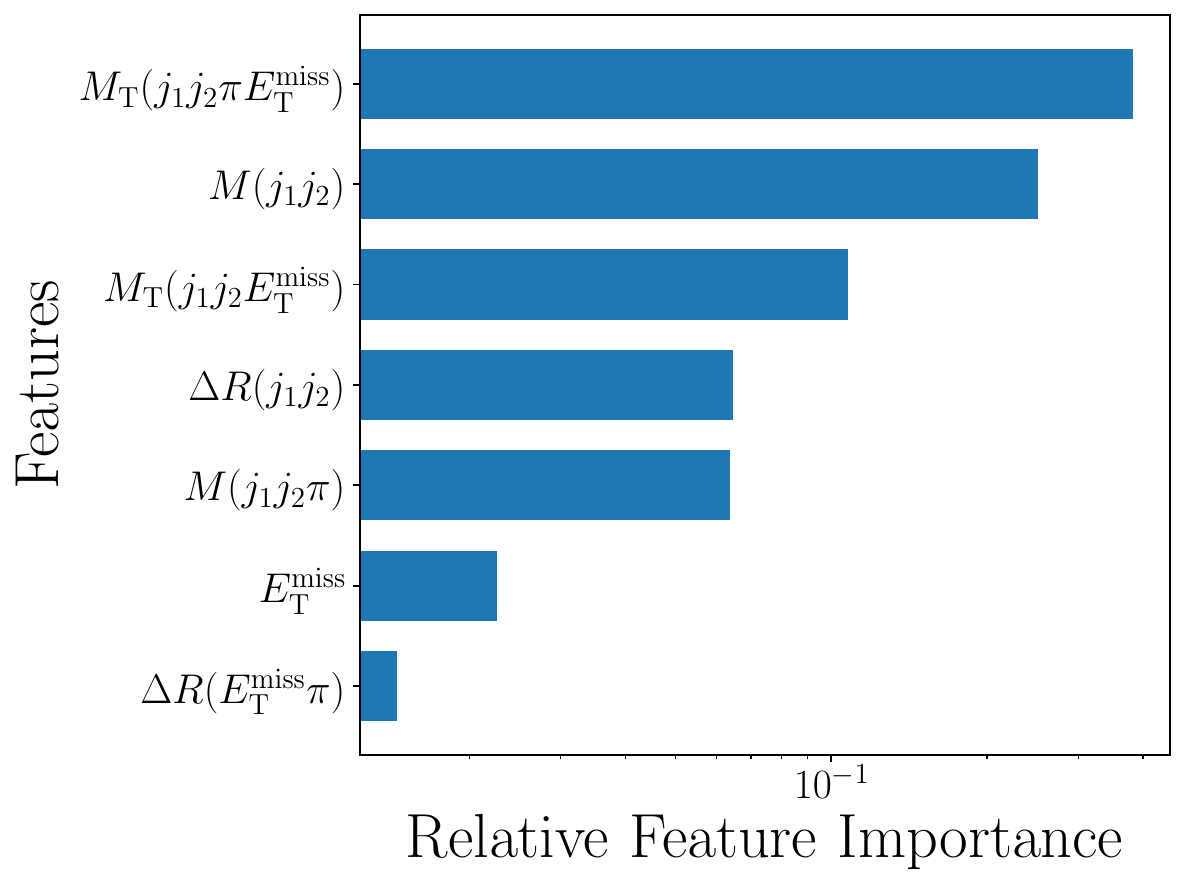}\hfill
    \includegraphics[width=0.32\textwidth]{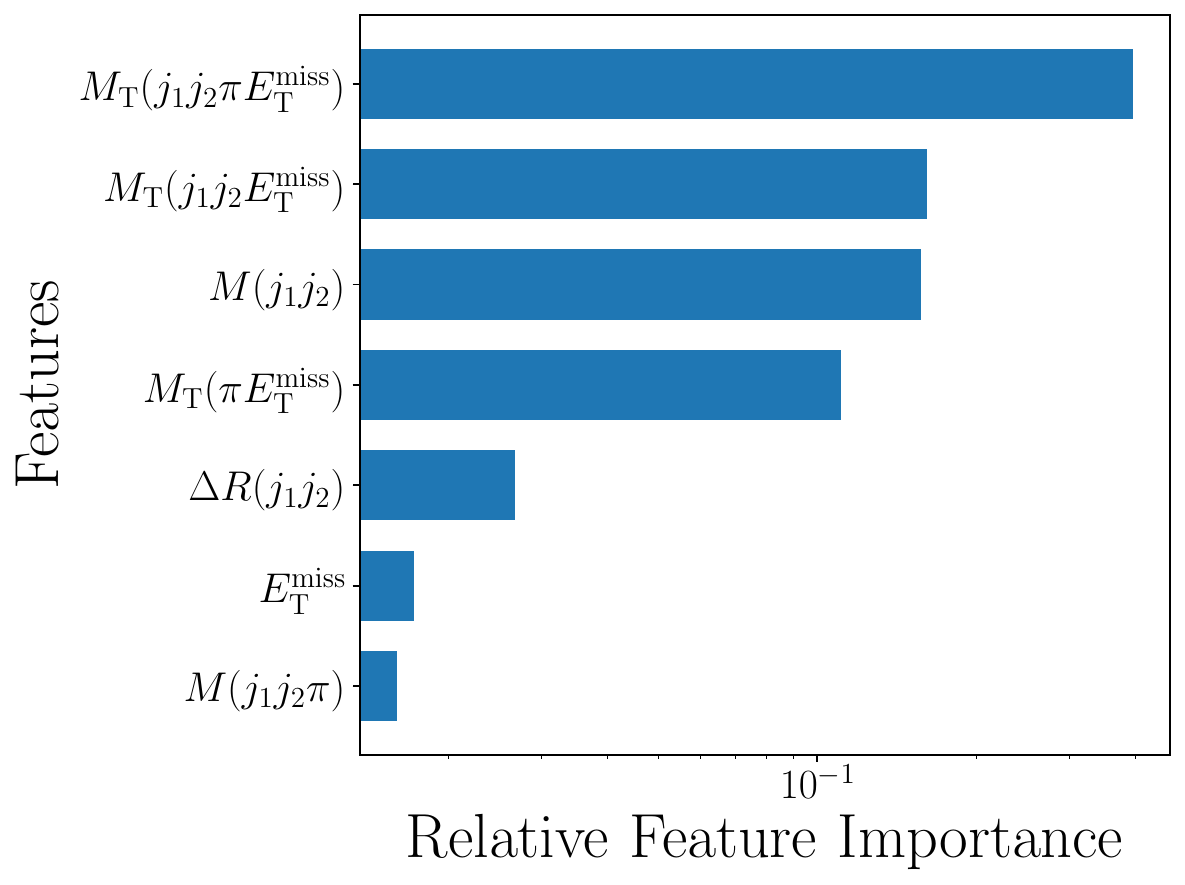}\hfill
    \includegraphics[width=0.32\textwidth]{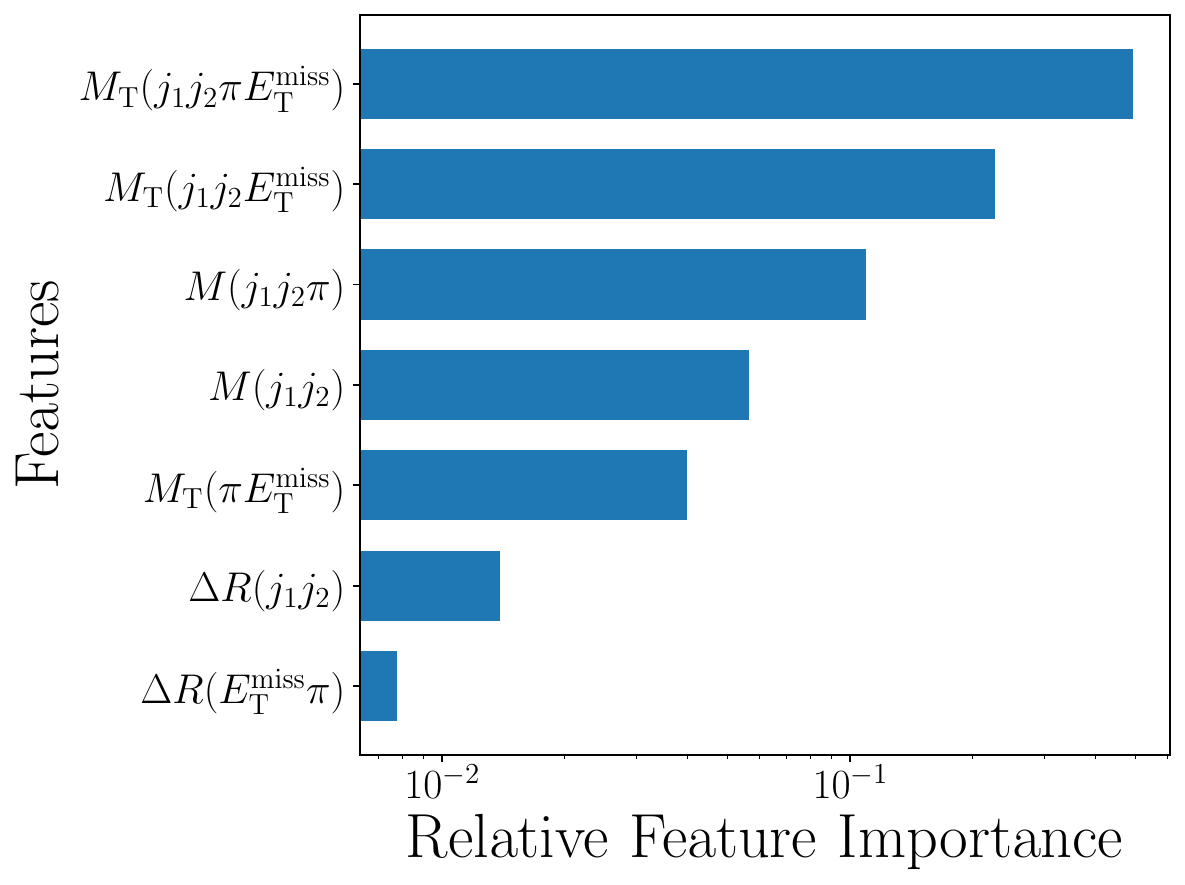}
    \caption{Relative feature importance for the BDT classifier at a heavy neutrino mass of 0.1 TeV (left), 1 TeV (middle), and 10 TeV (right) for $\ell = \tau$.}
    \label{fig: BDT_Relative_tau}
\end{figure*}

For the case of $\ell = \tau$, we focus exclusively on hadronic $\tau$ decays, given the reconstruction challenges associated with purely leptonic $\tau$ decays at the LHC and the improved performance of $\tau$-jet identification enabled by modern machine-learning techniques~\cite{CMS:2018jrd,CMS:2022prd,ATLAS:2022aip,CMS:2025kgf}. To model this scenario--and noting that hadronic $\tau$ decays produce several possible final states with different pion multiplicities--we use the TauDecay~\cite{Hagiwara:2012vz} package to simulate only the single charged-pion decay mode, $pp \to \tau(\to \pi \nu_\tau)\nu_\tau jj$, as a qualitative representation of hadronic $\tau$ decays for both the signal and background, applying the same selection criteria as before. During model training, the charged lepton is replaced by a charged pion, and the missing energy includes contributions from two light neutrinos. The resulting BDT output and relative feature importance are shown in FIG.~\ref{fig: BDT_tau} and FIG.~\ref{fig: BDT_Relative_tau}, respectively.

For similar reasons, the transverse mass between the charged pion and the missing energy also peaks around the $W$ boson mass for SM backgrounds, making it one of the most important features, as confirmed by FIG.~\ref{fig: BDT_Relative_tau}. However, because hadronic $\tau$ decays involve additional neutrinos in the final state, the kinematic reconstruction is less precise, which degrades the overall discriminating power of the input variables. As a result, the BDT model efficiency in the $\tau$ channel is reduced compared to the light lepton channels.

We also checked the possible interference between the heavy-neutrino signal and the SM $pp\to \ell\nu jj$ background. Such effects can arise from the subset of SM diagrams with the same external states as the signal, for example, when a light neutrino replaces the heavy neutrino. The signal (SIG), background (BKG), and SM--BSM interference (INF) samples are generated separately, normalized to the same integrated luminosity, and combined bin by bin. We quantify the interference through the relative magnitude $\lvert\mathrm{INF}\rvert/(\mathrm{SIG}+\mathrm{BKG})$ for representative benchmarks $m_N=0.1$, $1$, and $10~\mathrm{TeV}$. For the relevant kinematic observables which enter the BDT, e.g., $M_{\mathrm{T}}(\ell E_{\mathrm{T}}^{\mathrm{miss}})$, $M(j_1j_2)$, and $p_{\mathrm{T}}(\ell)$, the interference contribution is generally small. For $m_N=0.1$ and $10~\mathrm{TeV}$, the ratio is predominantly at the $10^{-3}$ level or below, while the $m_N=1~\mathrm{TeV}$ benchmark exhibits the largest effects, reaching the percent level in a few bins of the high-end tails of the distributions. We find that no representative distribution shows an interference contribution comparable to the uncertainties considered in our studies. The signal--background interference remains subleading and for all practical purposes does not modify the relevant kinematic distributions or the projected sensitivity. The small size of the interference can be understood qualitatively from the structure of the signal and background amplitudes. Only SM subprocesses with identical external partonic states and compatible color structures can interfere with the purely electroweak heavy-neutrino signal. In particular, many background contributions have color structures that are orthogonal to the color-singlet signal and therefore do not contribute after the color sum. In addition, the signal and SM background are enhanced in different regions of phase space due to their distinct resonant and production topologies, which limits their overlap and hence the size of the interference term. The interference can reach the percent level at intermediate $m_N$, where the signal transitions between the low-mass resonant regime and the high-mass VBF regime and has greater kinematic overlap with the SM background. At large $m_N$, the increasingly hard and VBF-like signal kinematics further separate the signal from the dominant SM background, while at low $m_N$ the resonant signal topology provides a more localized contribution.

A shape-based, binned profile likelihood analysis of the BDT score distributions is performed using the CMS \verb|Combine| tool~\cite{CMS:2024onh}. Expected limits on the signal production cross sections are obtained with the asymptotic method using an Asimov dataset. All results are presented assuming an integrated luminosity of $\mathcal{L}_{\mathrm{int}} = 3000~\mathrm{fb}^{-1}$, representative of the HL-LHC dataset. For the limit-setting procedure, the BDT output distributions (50 bins between 0--1) are scaled according to $N = \mathcal{L}_{\mathrm{int}} \times \sigma \times \epsilon$, where $\epsilon$ encapsulates event selection and reconstruction efficiencies (for the $\tau$ channel, the hadronic $\tau$ decay branching ratio is included in $\epsilon$).

Systematic effects are incorporated directly into the likelihood through nuisance parameters. Lognormal priors are assigned to normalization-related uncertainties, while Gaussian constraints are used for uncertainties affecting the shapes of the BDT output distributions. Both experimental and theoretical sources of systematic uncertainty are considered. To account for uncertainties associated with the choice of parton distribution functions, a normalization uncertainty in the range of 1--5\%, depending on the Monte Carlo sample, is applied following the PDF4LHC prescriptions~\cite{Butterworth:2015oua}. This contribution has a limited impact on the total expected yields and introduces negligible distortions to the BDT output shapes when compared to the statistical fluctuations in individual bins. PDF uncertainties are treated as uncorrelated between signal and background processes, but fully correlated across the BDT bins within a given process.

Additional theoretical uncertainties arising from missing higher-order corrections in the signal cross section are evaluated by varying the renormalization and factorization scales independently by factors of two. The resulting variations in the bin-by-bin BDT yields are found to be at the level of 2--5\%, depending on the heavy neutrino mass, mixing parameters, and the position within the BDT distribution.

Experimental uncertainties are implemented following current CMS performance studies. A conservative 3\% uncertainty on the integrated luminosity measurement is included~\cite{CMSBRIL:2022qin} and treated as fully correlated across all processes and BDT bins. For electrons and muons, a 2\% uncertainty is assigned to reconstruction, identification, and isolation efficiencies, along with an additional 3\% uncertainty associated with momentum and energy scale and resolution effects, assumed to be independent of lepton transverse momentum and pseudorapidity~\cite{CMS:2016fxb,CMS:2018iye,CMS:2017xcw,CMS:2024nmz,Avila:2018sja,CMS:2016xbv,Florez:2016lwi,CMS:2013pbl,CMS:2012lkx,Florez:2023jdb}. These lepton-related uncertainties are taken as correlated across all processes containing prompt leptons and across the BDT bins for each process.

Jet energy scale uncertainties, ranging between 2\% and 5\% depending on jet $p_\mathrm{T}$ and $|\eta|$~\cite{CMS:2016fxb,CMS:2018iye,CMS:2020wzx}, are propagated to the BDT output and induce shape variations of approximately 1--3\% across the discriminant bins. While a data-driven calibration of SM background predictions is beyond the scope of this study, an additional normalization uncertainty of 10\% is assigned to both signal and background expectations to reflect residual modeling uncertainties. For example, we note that the QCD multijet background is a negligible contribution in these studies, however, it is also notoriously difficult to model accurately using simulation because its contribution depends sensitively on detector effects, particularly the tails of the jet-energy response that generate fake missing transverse momentum. A precise estimate for an HL-LHC analysis will therefore require calibration using data-driven methods once collision data become available. Such an experimental treatment is beyond the scope of the present phenomenological study, however, the additional 10\% uncertainty on the background normalization is meant to cover such an effect and other residual background mismodeling uncertainties. This contribution is taken to be uncorrelated between signal and background processes and independent of the BDT output bin. When all systematic effects are combined, the resulting overall impact is conservatively estimated to be of order 20\%.

Finally, statistical uncertainties associated with the finite size of the simulated signal and background samples are included on a bin-by-bin basis in the BDT output histograms. These uncertainties range from approximately 1\% to 35\%, depending on the process and the specific region of the BDT discriminant, and are treated as uncorrelated across both processes and bins.

\begin{figure}[htbp]
\begin{center} 
  \includegraphics[width=0.45\textwidth, keepaspectratio]{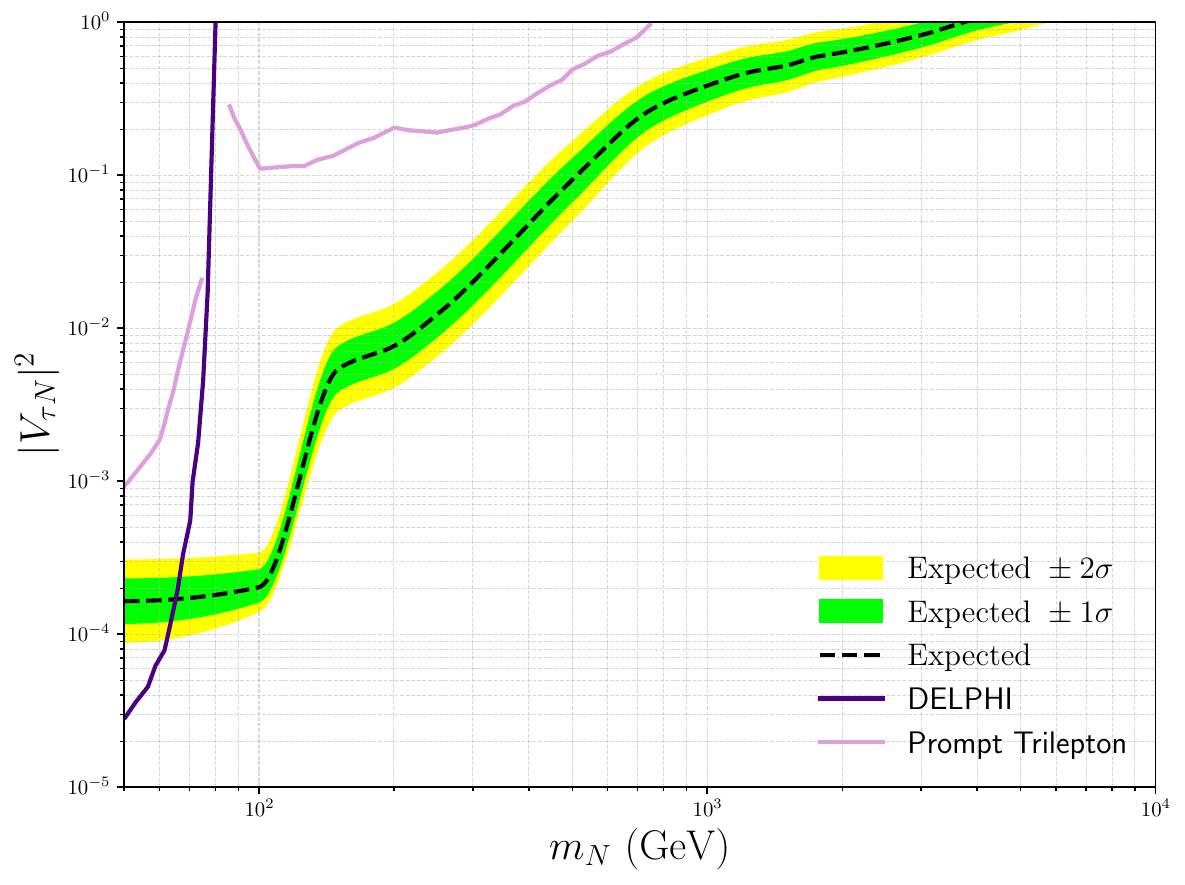}
\end{center}
\caption{Expected upper limits at 95\% confidence level on $\abs{V_{\tau N}}^2$ as a function of the heavy neutrino mass $m_N$. The solid line indicates the median expected limit, while the green and yellow bands represent the $\pm 1\sigma$ and $\pm 2\sigma$ expected uncertainty intervals, respectively. Results from the DELPHI~\cite{DELPHI:1996qcc} and prompt trilepton~\cite{CMS:2024xdq} searches are included for comparison.}
\label{fig: VtaN}
\end{figure}

\begin{figure}[htbp]
\begin{center} 
  \includegraphics[width=0.45\textwidth, keepaspectratio]{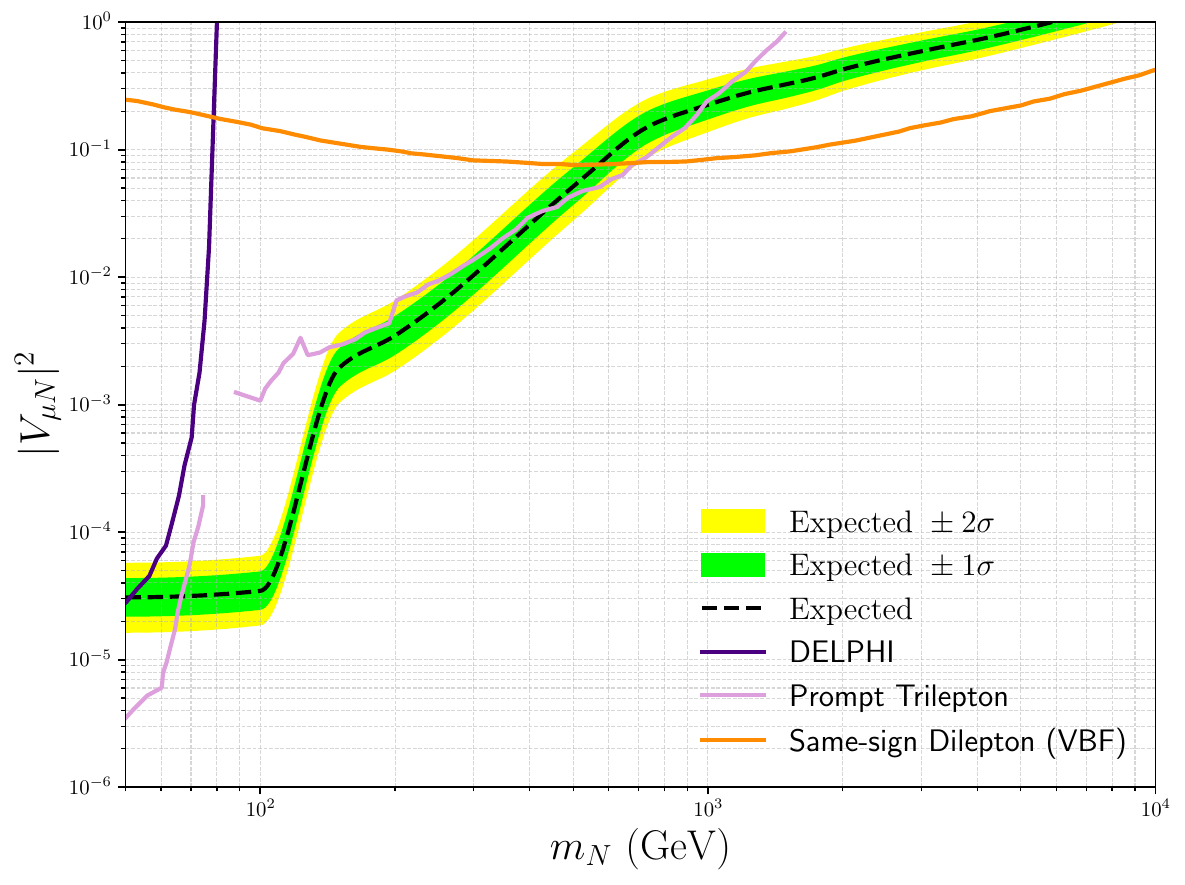}
\end{center}
\caption{Expected upper limits at 95\% confidence level on $\abs{V_{\mu N}}^2$ as a function of the heavy neutrino mass $m_N$. The solid line indicates the median expected limit, while the green and yellow bands represent the $\pm 1\sigma$ and $\pm 2\sigma$ expected uncertainty intervals, respectively. Results from the DELPHI~\cite{DELPHI:1996qcc} and prompt trilepton~\cite{CMS:2024xdq} searches are included for comparison.}
\label{fig: VmuN}
\end{figure}

Because the relationship between the cross section and the mixing parameter varies substantially across the mass range $m_N \in [50~\mathrm{GeV}, 10~\mathrm{TeV}]$, we generate additional samples for different mixing parameter values within $\abs{V_{\ell N}}^2 \in [0, 1]$ and fit the resulting cross sections using the parameterization
\begin{equation}
    \sigma = a \abs{V_{\ell N}}^2 + b \abs{V_{\ell N}}^4,
\end{equation}
where $a$ and $b$ are free parameters determined independently at each mass point. This fitted relation enables us to extract the sensitivity to the mixing parameter $\abs{V_{\ell N}}^2$ at the HL-LHC with an integrated luminosity of $3000~\mathrm{fb}^{-1}$. The results for $\ell = \tau$ and $\ell = \mu$ are shown in FIG.~\ref{fig: VtaN} and FIG.~\ref{fig: VmuN}, respectively.

\begin{figure}[htbp]
\begin{center} 
  \includegraphics[width=0.45\textwidth, keepaspectratio]{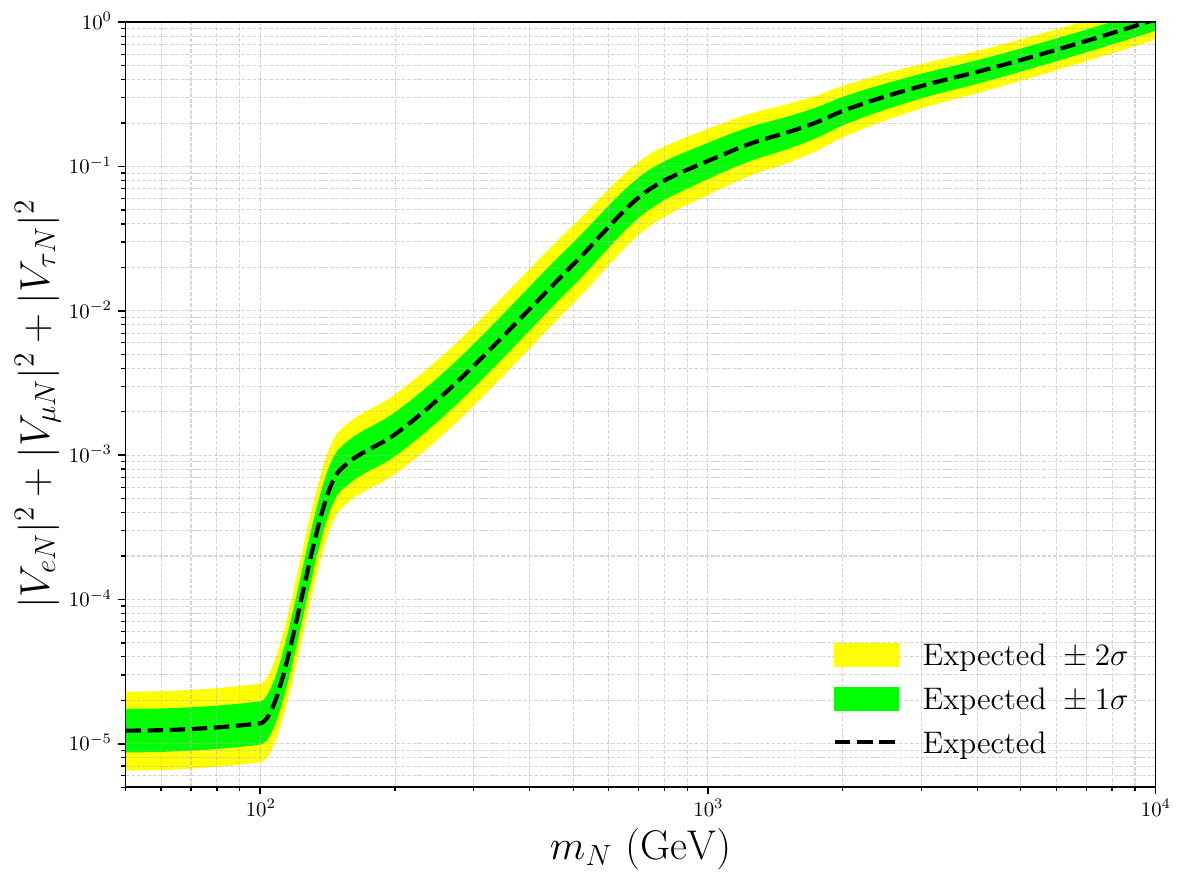}
\end{center}
\caption{Expected upper limits at 95\% confidence level on $\abs{V_{e N}}^2 + \abs{V_{\mu N}}^2 + \abs{V_{\tau N}}^2$ as a function of the heavy neutrino mass $m_N$. The solid line indicates the median expected limit, while the green and yellow bands represent the $\pm 1\sigma$ and $\pm 2\sigma$ expected uncertainty intervals, respectively.}
\label{fig: VeNVmuNVtaN}
\end{figure}

Since the flavor of the final-state light neutrino cannot be directly identified in collider experiments, we consider combining charged-lepton and neutrino flavors to probe possible violations of lepton universality. For $\ell = e, \mu$, this is implemented by replacing $\lvert V_{\mu N} \rvert^2$ with $\lvert V_{e N} \rvert^2 + \lvert V_{\mu N} \rvert^2$ and combining the two channels when computing the expected limits and uncertainties. This procedure can be straightforwardly generalized to include the $\tau$ lepton. Since the $\ell = e, \mu$ case is a subset of the $\ell = e, \mu, \tau$ combination and leads to qualitatively similar results, we present only the latter in Fig.~\ref{fig: VeNVmuNVtaN}.

Across FIG.~\ref{fig: VtaN} through FIG.~\ref{fig: VeNVmuNVtaN}, the sensitivity to the mixing parameter $\abs{V_{\ell N}}^2$ reaches values as low as $\order{10^{-5}}$ for a heavy neutrino mass of $m_N = 50~\mathrm{GeV}$, extending beyond the reach of previous searches such as those performed by DELPHI~\cite{DELPHI:1996qcc}. The sensitivity degrades rapidly as $m_N$ increases toward the TeV scale, and then decreases more gradually as VBF production becomes dominant. We emphasize that the upper boundary $\abs{V_{\ell N}}^2 \leq 1$ used in Figs.~\ref{fig: VtaN}--\ref{fig: VeNVmuNVtaN} should be interpreted only as the formal limit of the simplified mixing parametrization. The large-mixing region is not generically viable in minimal type-I seesaw scenarios, where the smallness of the observed light-neutrino masses, together with electroweak precision data, flavour measurements, neutrino-oscillation data, and existing direct searches, can impose strong additional constraints. Nevertheless, sizable active--sterile mixing can be realized in low-scale seesaw constructions, such as inverse or linear seesaw models, or in scenarios with an approximate lepton-number symmetry. Our results should therefore be viewed primarily as model-independent HL-LHC sensitivities to the $\ell\nu jj$ signal rate; the conversion of this rate into limits on $\lvert V_{\ell N}\rvert^2$ is model-dependent and should be interpreted within a specified ultraviolet completion. In particular, we demonstrate that a novel VBF production mechanism mediated by virtual $t$ channel heavy neutrino exchange, when combined with advanced machine-learning--based analysis techniques, can lead to a substantial enhancement in sensitivity to third-generation heavy neutrino searches involving tau leptons--an experimentally challenging regime where current constraints remain weak. For light-lepton final states, the proposed strategy can extend the discovery reach at the HL-LHC and offers a complementary probe to existing LHC search methodologies.

We close this section by noting that different choices of $V_{\ell N}$ can modify the kinematic distributions, thereby affecting the training of the BDT models and the final projected sensitivity to $\abs{V_{\ell N}}^2$. In the low-mass regime, the heavy neutrino can be produced resonantly, and the $s$-channel contribution is always dominant due to the scaling relation between $\abs{V_{\ell N}}^2$ and the total cross section. For larger masses, if $\abs{V_{\ell N}}^2$ is scaled down to very small values, the resulting cross section becomes too suppressed, placing $\abs{V_{\ell N}}^2$ outside the projected sensitivity of our analysis. This consideration, therefore, justifies the range of $\abs{V_{\ell N}}^2$ values presented in our results.

\section{Conclusion}

In this study, we have proposed and demonstrated a potential strategy for probing heavy neutrinos at the LHC using $\ell \nu jj$ final states analyzed with machine-learning techniques. We simulated signal and SM background events over a broad heavy neutrino mass range from 50 GeV to 10 TeV, incorporating both $s$ channel production--dominant at low masses--and VBF processes which become increasingly important at higher masses and maintain sensitivity in the multi-TeV regime.

By employing modern machine-learning methods, we trained BDT models on an extensive set of kinematic features, achieving strong discrimination between signal and background. The feature-importance analysis shows that transverse masses involving missing transverse energy and charged leptons provide key discriminating power. We also extended the analysis to hadronic tau decays to assess the sensitivity to the mixing parameter $\abs{V_{\tau N}}^2$.

Using the CMS \verb|Combine| tool, we obtained expected upper limits on the mixing parameter $\abs{V_{\ell N}}^2$ at each mass point, assuming an integrated luminosity of $3000~\mathrm{fb}^{-1}$ at the HL-LHC. The projected sensitivity reaches $\order{10^{-5}}$ for $m_N = 50~\mathrm{GeV}$ and remains competitive up to the multi-TeV scale. We further explored scenarios where contributions from different charged-lepton flavors are combined, enabling tests of potential lepton-universality violation and providing a more inclusive framework for heavy neutrino searches.

In conclusion, an analysis strategy that combines inclusive heavy neutrino production—encompassing both $s$ channel and vector boson fusion mechanisms—with a machine-learning-based search significantly extends the discovery reach at hadron colliders. This approach simultaneously improves sensitivity in both the low- and high-mass regimes and markedly improves coverage of third-generation heavy neutrinos involving $\tau$ leptons, an experimentally challenging sector where existing constraints remain comparatively weak.
This methodology complements existing search strategies and highlights promising opportunities to explore physics beyond the SM in the neutrino sector.

\begin{acknowledgements}

We thank Umar Sohail Qureshi for discussions on the application of machine learning techniques. We gratefully acknowledge support from Vanderbilt University and the U.S. National Science Foundation. This work is supported in part by NSF Awards PHY-1945366 and PHY-2411502. AF thanks the constant and enduring financial support received for this project from the faculty of science at Universidad de Los Andes (Bogot\'a, Colombia).

\end{acknowledgements}

\noindent\textbf{Data Availability Statement.}
The datasets generated during and/or analysed during
the current study are available from the corresponding
author on reasonable request.

\medskip
\noindent\textbf{Code Availability Statement.}
The code/software generated during and/or analysed
during the current study is available from the
corresponding author on reasonable request.

\end{document}